%% file: paper_body.tex
\documentclass[aps,prx,showpacs,preprintnumbers,twocolumn,superscriptaddress,10pt]{revtex4-2} 
\usepackage{packages}
\newcommand{\id}{\mathds{1}}
\newcommand{\Stot}{S_{\mathrm{tot}}}
\newcommand{\Stota}{\langle \Stot \rangle}
\newcommand{\pc}{p_{\mathrm{c}}}
\newcommand{\ps}{\ket{0}^{\otimes N}}
\newtheorem{theorem}{Theorem}
\newcommand{\mpipks}{Max Planck Institute for the Physics of Complex Systems, N\"othnitzer Str.\ 38, 01187 Dresden, Germany}
\newcommand{\julich}{Institute of Quantum Control (PGI-8), Forschungszentrum J\"ulich, D-52425 J\"ulich, Germany}
\newcommand{\regensburg}{University of Regensburg, Universit\"atsstr.\ 31, Regensburg D-93053, Germany}

\begin{document}

\author{Giovanni Cemin}
\email{gcemin@pks.mpg.de}
\affiliation{\mpipks}

\author{Markus Schmitt}
\thanks{equal contribution}
\affiliation{\regensburg}
\affiliation{\julich}

\author{Marin Bukov}
\thanks{equal contribution}
\affiliation{\mpipks}

\clearpage
\title{
Learning to stabilize nonequilibrium phases of matter \\
with active feedback using partial information
}

\begin{abstract}
    Recent advances in quantum computing experiments have enabled new pathways for feedback control. Here, we investigate the role of information in interactive quantum many-body dynamics to engineer out-of-equilibrium phases of matter.
    To this end, we train reinforcement learning agents to prevent entanglement spreading in $(1{+}1)$-dimensional stabilizer circuits with up to $128$ qubits.
    We find that, above an information threshold, agents learn stochastic near-optimal strategies that prevent the system from reaching volume-law entangled states and instead stabilize area-law scaling. 
    These strategies rely on bottlenecks that divide the system and cap the entanglement in the steady state.
    Crucially, learned strategies are inherently out of equilibrium, require real-time active feedback, and could not be reproduced by simple human-designed control rules.
    This work establishes the foundations for information-driven individual control of many interacting quantum degrees of freedom, demonstrating the capabilities of reinforcement learning to stabilize and uncover novel critical properties of many-body nonequilibrium steady states.
\end{abstract}

\maketitle

\date{\today}

\stoptoc
\section{Introduction}
\begin{figure*}[t!]
    \centering
    \includegraphics[width=\textwidth]{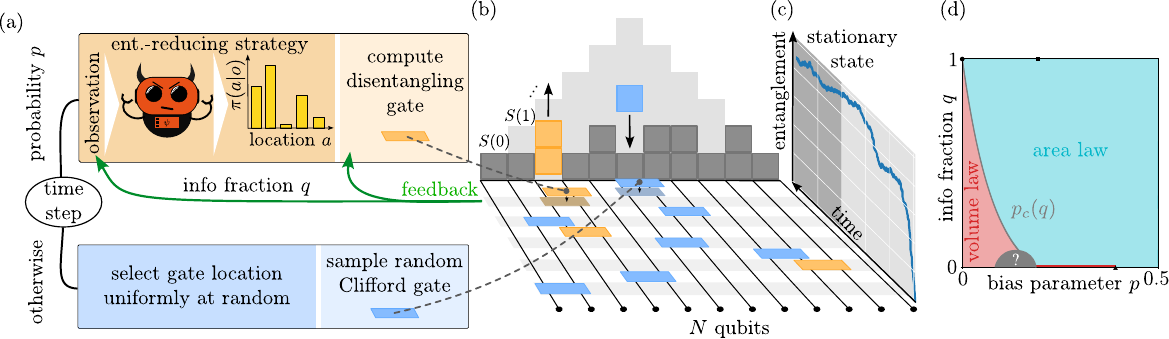}
    \caption{
    \textbf{Schematic representation of the stochastic $(1{+}1)$-dimensional stabilizer circuit dynamics and phase diagram.}
    A chain of $N$ qubits evolving under stabilizer circuit dynamics is initialized in the product state $\ket{0}^{\otimes N}$. 
    (a) At each time step, a two-qubit Clifford gate is applied via biased selection parameterized by $p$: feedback (green arrow) provides an observation $o$ containing a fraction $q$ of the system's state information, used by the \textit{entanglement-reducing strategy} $\pi(a|o)$ to select a location $a$ for placing optimally disentangling gates (orange rectangles); 
    the complementary process samples randomly both the location and gate (blue rectangles).
    (b) Entanglement entropy $S(a)$ across each bond is represented by blocks: random gates increase entanglement on average (adding blocks), whereas disentangling gates reduce it (removing blocks).
    (c) Competition between entanglement-reducing strategy and stochastic dynamics leads to a non-equilibrium steady state (darker shaded region). 
    (d) Average steady-state entanglement for active-feedback strategies exhibits area-law phases when strategies have sufficient information ($q$ large) or strong disentangling bias ($p$ large). The phase diagram (schematic) of information fraction $q$ vs.~bias parameter $p$ exhibits a phase transition between volume- and area-law phases, at a critical information threshold $\pc(q)$.
    The grey region indicates where finite-size effects are too strong to draw quantitative conclusions.
    }
    \label{fig:model}
\end{figure*}
Information, as conceptualized by Shannon, constitutes a universal framework spanning both classical and quantum physics. 
Its origins trace back to Boltzmann's statistical formulation of thermodynamics, where entropy quantifies uncertainty and the lack of information about a system's microscopic state.
An instance of the intricate interplay between entropy and information is Maxwell's demon, which reduces the entropy of a finite-temperature gas by controlling the dynamics of its many particles while keeping a record of their positions and momenta.
This seeming violation of the second law of thermodynamics was resolved by Landauer: the demon's finite memory requires erasing the information on its record -- an irreversible process that generates entropy~\cite{landauer_1961}.
Despite significant progress, our understanding of the role of information in controlling many-body states remains rudimentary. 
In parallel, experiments emulating Maxwell's demon have become feasible in quantum systems using feedforward control~\cite{salathe_2018,sakaguchi_2023,iqbal_2024,buhrman_2024}. Thus, the question arises of how to extract and utilize information to manipulate quantum many-body states. 

Quantum information, however, is inextricably tied to entanglement~\cite{wendin_2017,preskill_2022,gross_2023}. 
Controlling entanglement between qubits is instrumental for implementing algorithms that offer quantum advantage~\cite{biamonte_2020,daley_2022,fux_2024,zhao_2025}, making it a central challenge in quantum technology~\cite{aolita_2015,bauer_2020,diez_2021,niedermeier_2024,nakhl_2024}.
Key achievements include the preparation of many-body entangled states~\cite{mooney_2021,zhang_2022,zhu_2022,chen_2023} and the implementation of protocols to measure entanglement in experiments~\cite{islam_2015,vermersch_2018,brydges_2019}.
Theoretically, our knowledge of many-body entanglement has benefited from improved understanding of unitary quantum circuits~\cite{vasseur_2016,nahum_2018,swingle_2018,li_2019,fisher_2023,sierant_2023}. 
Unitary dynamics interspersed with quantum measurements can lead to entanglement transitions between area- and volume-law phases of the nonequilibrium steady state~\cite{li_2018,skinner_2019,chan_2019,sierant_2022,sierant_2023_control}. 
This critical phenomenon emerges from a competition: unitary evolution generates and spreads entanglement throughout the system, while measurements destroy entanglement by extracting classical information.
However, experimental realizations of measurement-induced phase transitions are obstructed by a postselection bottleneck~\cite{noel_2022,hoke_2023,koh_2023}.

This difficulty is alleviated in \textit{unitary circuit games}, which exhibit entanglement transitions occurring in purely unitary dynamics~\cite{nahum_2017,yepes_2024,yepes_2025}. 
At each step of the game, a coin toss with fixed bias $p$ determines which of two agents acts: one applies a disentangling gate; the other, a random local unitary that produces volume-law entanglement in the circuit. Gates are placed at random locations. 
Varying $p$ biases the competition between the agents, driving the system into different nonequilibrium steady states.
In the thermodynamic limit, a transition between area- and volume-law entangled phases occurs: Clifford circuits feature a finite critical bias $\pc$, whereas Haar-random circuits exhibit only a volume-law phase~\cite{yepes_2024}.
Whereas information about the quantum state is used to determine the disentangling gate, it is ignored when selecting the gate location. This severely constrains both the controllability of entanglement and the accessible critical properties of the steady state. \\ 

Here, we raise the question about the role of information in the active feedback control of quantum many-body states. Leveraging partial state information to optimally manipulate quantum dynamics defines a broad framework applicable across emerging near-term quantum computing devices, where quantum states are not fully observable. 
Focusing on entanglement-reducing strategies for unitary circuit games, we introduce feedback-driven control that utilizes partial quantum state information.
We demonstrate that access to information severely transforms the physics, allowing us to uncover emergent patterns in the control protocol and engineer new critical properties in the nonequilibrium steady state. 
We implement these strategies using reinforcement learning (RL)~\cite{sutton_2018,bukov2026reinforcement,bukov_2018,fosel_2018,niu_2019,wauters_2020,guo_2021,metz_2023,tashev2024,olle_2024}: akin to Maxwell's demon, RL agents are designed for active feedback~\cite{sordal_2019,rao_2023,erdman_2025}; unlike quantum optimal control algorithms, they can naturally handle partial information.
Our numerical simulations indicate that, above a critical information threshold, near-optimal informed strategies may be capable of completely inhibiting entanglement growth.

Specifically, we consider $(1{+}1)$-dimensional stabilizer circuits with up to $128$ qubits, and investigate random, greedy, and learned \textit{entanglement-reducing} strategies.
Greedy strategies outperform random ones, yet stochastic circuit dynamics render them suboptimal: they lack active feedback, essential for efficiently preventing entanglement spreading. 
Remarkably, all learned strategies achieve area-law steady states maintaining low entanglement even with rare disentangling gate placements; this is accomplished by introducing a sequence of bottlenecks that generate pyramidal structures in the long-time spatial entanglement distribution, effectively partitioning the qubit chain into subsystems and limiting the maximum entanglement that can be accessed. 
Moreover, we demonstrate that complete state information is not necessary for implementing efficient entanglement-reducing strategies. Below a critical information threshold, learned strategies fail, and the steady state reverts to volume-law entanglement. 
Finally, the human-designed strategies we tested were unable to reproduce the area-law steady state, which suggests the importance of active feedback.
Our study directly demonstrates the capabilities of the reinforcement learning framework to successfully control the dynamics of individual degrees of freedom in extensive quantum many-body systems.

\section{Model}
\label{sec:model}

We consider a $(1{+}1)$-dimensional stabilizer circuit where $N$ qubits are arranged in a one-dimensional chain and evolve in discrete time from an initial product state $\ket{\psi_0} = \ket{0}^{\otimes N}$. 
Stabilizer circuits are amenable to efficient classical simulation \cite{gottesman1996,Gottesman_1998,Gottesman_1998_Heisenberg,aaronson_2004} (see App.~\ref{app:stabilizer_circuits}), enabling the investigation of hundreds of qubits, otherwise prohibitive for generic quantum circuits.
The classical simulability arises because stabilizer states can be represented by a so-called tableau, an $N \times(2N+1)$ binary matrix that fully characterizes the quantum state.
Nonetheless, Clifford circuits exhibit rich quantum behavior, including entanglement growth~\cite{sang_2022,richter_2023}, phase transitions~\cite{lunt_2021,makki_2024}, and nontrivial dynamical phases~\cite{zhuang_2023}, with a computational complexity polynomial in the system size $N$.

The chain evolves according to simple rules, as shown in Fig.~\ref{fig:model}(a).
At each time step, the bias parameter $p$ determines the probability for the disentangling agent to act.
When selected, an optimally disentangling two-qubit gate is applied using an \textit{entanglement-reducing strategy} $\pi (a|o)$, where $o$ is the observation of the current state and $a$ denotes the bond location $a \in \{0,\dots, N-2\}$ (bonds counting from the left) where the gate is applied.
Otherwise, a two-qubit gate randomly sampled from the Clifford group is applied at a uniformly random location; while these random gates can occasionally disentangle, on average, they increase entanglement.
Thus, the bias parameter $p$ controls the balance between disentangling and stochastic evolution. 
In this framework, the strategy $\pi (a|o)$ accesses partial state information in the observation $o$ through feedback, implementing active information-driven control strategies beyond the passive dynamics of unitary circuit games~\cite{yepes_2024}.
At long times, the circuit reaches a non-equilibrium steady state (see App.~\ref{app:nonequilibrium}) whose properties depend on the strategy employed.

Reducing entanglement constitutes a coupled optimization problem over the gate locations and the structure of gates themselves.
Solving the full problem is complex and can result in black-box strategies that obscure underlying physical mechanisms.
Therefore, to retain interpretability, we decompose the problem into two steps: (i) the entanglement-reducing strategy that outputs the strategy $\pi(a|s)$ -- a probability distribution across the locations $a$ based on information about the current state $s$, and (ii) the application of an optimally disentangling gate at the selected location $a$ [Fig.~\ref{fig:model},(a) orange box]; for stabilizer states, we can efficiently identify such gates, as the reduced two-qubit density matrix is directly mapped to the appropriate disentangling Clifford gate (see App.~\ref{app:disentangling_cliffords}).
We feed back the instantaneous stabilizer tableau in the clipped gauge~\cite{nahum_2017,li_2019} to the entanglement-reducing strategy.
To model control with partial information, we introduce the \textit{information fraction} $q$, whereby rows of the tableau are randomly removed with rate $1-q$, resulting in a mixed-state description (see Sec.~\ref{sec:partial_info}).

At each time step, we compute the von Neumann entanglement entropy $S(a)$ across every bond $a=0,1,\dots, N-2$, which corresponds to all contiguous bipartitions of the chain.
Not all entanglement-reducing strategies that we consider are translationally invariant; hence, measuring entanglement at a single cut (e.g., at half-chain) is insufficient to characterize the system's global behavior.
To detect critical behavior in the long-time steady state, we define the \textit{total entanglement}
\begin{equation}
    S_\text{tot} := \sum_{a=0}^{N-2} S(a)
    \label{eq:Stot}
\end{equation}
as a global order parameter to detect entanglement phase transitions.
We study the normalized total entanglement $\Stota / \mathcal{N}$, where the normalization factor is the maximum value $\Stot$ can attain (cf.~App.~\ref{app:rl_environment})
\begin{equation}
    \mathcal{N}(x) =
        \left( \left\lfloor \dfrac{x}{2} \right\rfloor + 1 \right)^2, \qquad \text{$x$ odd integer}\, .
    \label{eq:normalization}
\end{equation}
Here, $x$ denotes the number of bonds in the chain, that is $x=N-1$.

In stabilizer circuits, the entanglement entropy is quantized: $S(a){=}n(a) \ln 2$, where $n(a) = \{0, 1, 2, \dots, \min(a + 1, N - 1 - a) \}$ is a bond-dependent integer, constrained by the local Hilbert space dimension (cf.~App.~\ref{app:stabilizer_circuits}). Moreover, considering contiguous bipartitions imposes a local constraint on the entropy profile: $-1 \leq S(a) - S(a+1) \leq 1$.
The discrete structure of entanglement in Clifford circuits allows for a natural and intuitive graphical representation, as illustrated in Fig.~\ref{fig:model}(b), where the total block height above each location $a$ encodes the value of the local entropy $S(a)$.
The constraint ensures that adjacent blocks differ in height by at most one unit.

To measure the long-time value $\langle \Stot \rangle$, we wait until the system reaches the steady state and then average over time and ensembles. 
We verify that the system has reached steady state using the Geweke Z-test \cite{geweke_1992}.
The required equilibration time and number of samples depend on the strategy, the bias $p$, and the system size $N$ (larger systems are effectively self-averaging).  Full details of the sampling methodology can be found in App.~\ref{app:sampling_steady_state}.\\

\section{Human designed strategies}

The model introduced above provides an ideal testbed for exploring entanglement-reducing strategies in many-body quantum systems. 
Starting from the initial product state $\ps$, bare stochastic dynamics ($p=0$) drives the system into an equilibrium state with maximal (volume-law) entanglement.
For $p>0$, the dynamics are partially controlled, driving the system to (non)equilibrium steady states (see App.~\ref{app:nonequilibrium}) with finite average entanglement $0 \leq \Stota / \mathcal{N} \leq 1$.
Unlike equilibrium steady states, nonequilibrium conditions offer greater flexibility in shaping the entanglement of the steady state.
At equilibrium, a closed chaotic circuit relaxes to an effectively infinite-temperature state with maximal entanglement~\cite{page_1993}, whereas area-law scaling is essentially tied to gapped ground states~\cite{hastings_2007}.
Out of equilibrium, the steady-state entanglement can be tuned continuously, as we will demonstrate below.
More broadly nonequilibrium conditions grant access to phases and critical behavior without equilibrium counterparts, such as measurement-induced transitions~\cite{li_2018,skinner_2019,fisher_2023}, many-body localization~\cite{nandkishore_2015}, Floquet and time-crystalline phases~\cite{khemani_2016,else_2016,zaletel_2023}, and driven-dissipative criticality~\cite{diehl_2008,sieberer_2016}.

For bias $p\ge 0.5$, the optimal entanglement-reducing strategy is trivial: place a disentangling gate at the same location where an entangling gate was previously placed.
This is allowed because disentangling gates are placed more (or equally) frequently than random gates.
In the regime $p<0.5$, disentangling gates are less frequent than random ones, and the stochastic dynamics acts as a strong entangling drive.
Maintaining the system out of equilibrium and achieving area-law entanglement or volume-law states with finite $\Stota$ requires effective, information-driven control strategies that can actively suppress entanglement growth.

\begin{figure}[t!]
    \includegraphics[width=\linewidth]{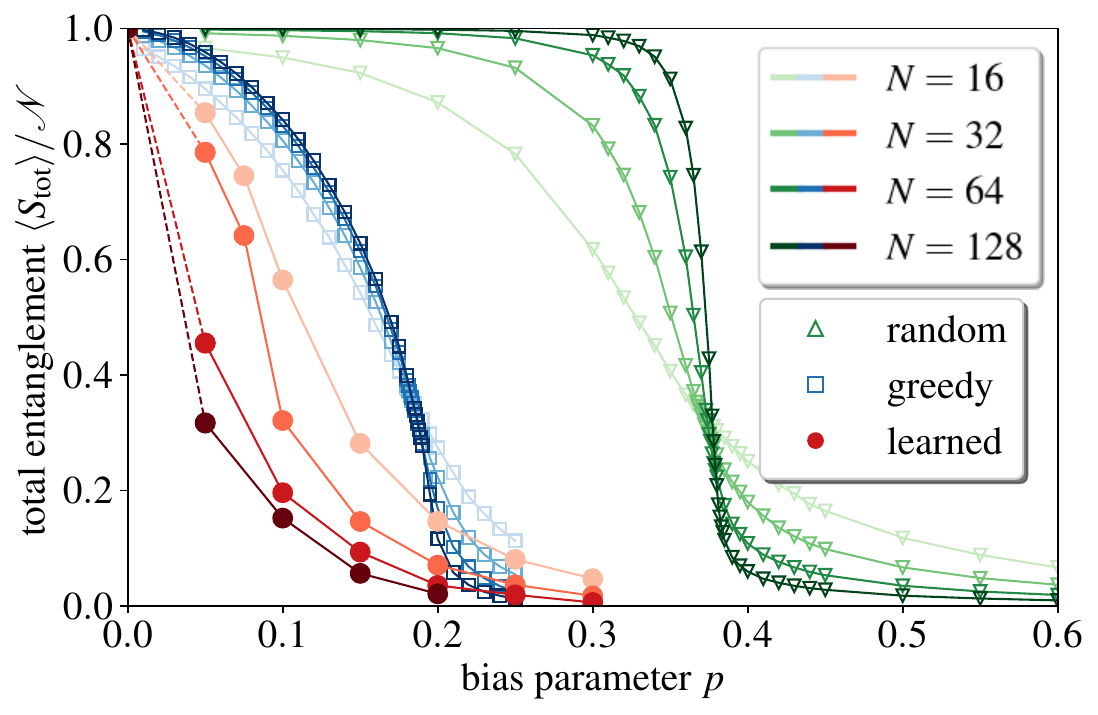}
    \caption{
    \textbf{Normalized ensemble-averaged steady-state entanglement $\langle \Stot \rangle$ vs. bias parameter $p$.}
    The random strategy data (green, triangles) indicate a
    transition between a nonequilibrium area-law and an equilibrium volume-law steady state at $\pc^\mathrm{random}{=} 0.384 \pm 0.003$, as $N {\rightarrow} \infty$.
    The greedy strategy (blue data, squares) exhibits a qualitatively different transition between two nonequilibrium steady states at $\pc^\mathrm{greedy}{=} 0.195 \pm 0.008$: the volume-law phase features a finite average total entanglement $0 {<} \Stota/\mathcal{N} {<} 1$, as $N{ \rightarrow} \infty$.
    By contrast, the learned strategy (red, circles) results in a nonequilibrium area-law steady state for all $p$ studied
    (see Sec.~\ref{sec:informed_strategies}).
    Sampling methodology and supplementary datasets can be found in Appendix Sec.~\ref{app:supp_results}.
    }
    \label{fig:combined}
\end{figure}

\subsection{Random strategy}
\label{sec:random}
As a baseline, the simplest entanglement-reducing strategy
selects gate locations $a$ uniformly at random, without state information or feedback, while the applied gates are still optimally disentangling. 
We refer to this as the random strategy, defined by
\begin{equation}
    \pi_\mathrm{random}(a|s) \equiv \pi_\mathrm{random}(a) = \frac{1}{N-1}.
    \label{eq:random_policy}
\end{equation}
In Fig.~\ref{fig:combined}, the green curves (triangles) show the steady-state entanglement $\Stota/\mathcal{N}$ as a function of the bias parameter $p$ for different sizes $N$. 
The data indicate a transition between an area-law and a volume-law entangled phase at $\pc^\mathrm{random}$, where the entanglement exhibits a discontinuous jump in the thermodynamic limit $N\to\infty$.
Finite-size scaling analysis (see App.~\ref{app:fss}) yields $\pc^\mathrm{random} =  0.384 \pm 0.003$~\footnote{The critical value $\pc$ is smaller than $0.5$ due to an asymmetry in the gates. The entanglement-reducing strategy exclusively places gates that reduce entanglement, while the random strategy samples gates uniformly from the Clifford group $\mathcal{C}_2$ -- most of which increase entanglement, but some of which decrease it. This asymmetry creates a slight bias favoring the entanglement-reducing strategy, resulting in $\pc < 0.5$.}, consistent with Ref.~\cite{yepes_2024}. 
The system reaches a non-equilibrium steady state with area-law entanglement for $p>\pc^\mathrm{random}$, and volume-law entanglement for $p<\pc^\mathrm{random}$. 
The non-equilibrium nature of the steady state arises from the violation of detailed balance due to information-based control, which introduces biased state transitions and entropy production. 
In the volume-law regime, biased transitions vanish as $1/N$, and the steady state becomes equilibrium in the thermodynamic limit (see App.~\ref{app:nonequilibrium} for further discussion).
This behavior is similar to the transition between a thermal and a localized phase~\cite{nandkishore_2015}.

\subsection{Greedy strategy}
\label{sec:greedy}

The random strategy illustrates nontrivial behavior and phase transitions emerging from the competition between control and stochastic dynamics.
However, being state-agnostic with no internal structure, it does not exploit quantum state information. 
By contrast, the greedy strategy represents the most direct approach for leveraging complete state information to reduce entanglement.

When placing a gate, the greedy agent evaluates all possible gate locations and selects uniformly at random among locations $a$ such that $\Delta S (a) > 0$, where $\Delta S(a) := S_t(a) - S_{t+1}(a)$ is the change in entanglement due to the disentangling gate at location $a$. 
We refer to this as the greedy strategy, defined by
\begin{equation}
    \pi_{\mathrm{greedy}}(a|s) = 
    \begin{cases}
        \frac{1}{Z}, & \text{if } \; \Delta S(a) > 0 \\
        0, & \text{otherwise.}
    \end{cases}
    \label{eq:greedy_policy}
\end{equation}
where $Z = |\{a: \Delta S (a) > 0 \} |$ is the number of locations with nonzero $\Delta S$; the dependence of $\Delta S (a)$ on the state $s$ is implicit.

In Fig.~\ref{fig:combined}, the blue curves (squares) show the steady-state average entanglement $\langle \Stot \rangle / \mathcal{N}$ as a function of bias $p$ for different system sizes $N$.
Compared to the random strategy, the data reveal a quantitatively and qualitatively different steady state.
First, finite-size scaling analysis yields a significantly lower critical bias $\pc^\mathrm{greedy} = 0.195 \pm 0.008 < \pc^\mathrm{random}$ (see App.~\ref{app:fss})  -- marking the transition between area- and volume-law entanglement in the thermodynamic limit. 
The feedback that contains state information enables more effective entanglement reduction, requiring fewer disentangling gates to achieve area-law scaling.
By design, the greedy strategy always selects locations where entanglement is reduced, unlike random placement, which may target ineffective locations.
Second, the behavior below $\pc^\mathrm{greedy}$ differs qualitatively: the random strategy equilibrates for $p<\pc^\mathrm{greedy}$ as $N\rightarrow \infty$, whereas the greedy strategy sustains non-equilibrium steady states with a finite average total entanglement entropy $0 < \Stota / \mathcal{N} < 1$. 
Crucially, the greedy strategy possesses internal degrees of freedom, consumes energy, and exchanges information with the system. 
The feedback loop continuously generates entropy through information processing, violating detailed balance and sustaining nonequilibrium conditions (see App.~\ref{app:nonequilibrium} for details).

This behavior can be explained by examining the entanglement-profile dynamics in the volume-law phase, i.e., for $p < p_c^\mathrm{greedy}$ (see supplementary videos, App.~\ref{app:captions}). 
Figure~\ref{fig:rl_fullinfo}(a) (left, blue region) shows snapshots at $p=0.05,0.10,0.15$ and system size $N=128$. 
In the steady state, the disentangling gates concentrate in a subregion of $N'(p) < N$ sites at the center of the chain, where $N'(p)$ is an increasing function of $p$. 
Outside this subregion, the entanglement is maximal and is left unchanged by any gate action. Consequently, only gates acting within the subregion can decrease or increase entanglement. This reduces the probability that a randomly placed gate has any effect from $1-p$ to $(1-p)\, N'/N$, enabling efficient entanglement suppression.
By reducing entanglement in the subregion $N'(p)$, the agent suppresses entanglement growth and preserves a finite total entanglement value below $\pc$.
Crucially, access to state information allows the existence of non-equilibrium steady states across all $p>0$, demonstrating how feedback can yield fundamentally different steady states and dynamical phases of matter, compared to passive evolution. \\

\section{Learned strategies}
\label{sec:learned_str}

The greedy strategy, despite its simple and rigid structure, shows that access to state information can be actively exploited to alter the system’s dynamics and steady-state properties.
However, the greedy strategy is inherently myopic: it optimizes entanglement reduction locally in time, evaluating only the immediate effect of each action without considering future consequences. 
More sophisticated approaches evaluate action sequences over multiple future steps to account for temporal correlations and delayed effects (e.g., tree search, Monte Carlo planning~\cite{sutton_2018}); however, this becomes computationally prohibitive due to the exponential growth of the search space with the planning horizon.
This type of Markov decision process with long-term consequences is precisely where deep reinforcement learning (RL) excels. 

Deep RL methods learn effective long-term strategies through trial and error, without requiring explicit enumeration of all possible future trajectories, making them well-suited for the high-dimensional, sequential decision-making required in quantum many-body control.
Moreover, RL naturally handles partially observed Markov decision processes, where the agent has access to only partial information about the system's state. 
This allows us to systematically investigate the role of information by varying the degree of observability, probing whether complete state access is essential, or whether partial information suffices to learn effective entanglement-reducing strategies.

\subsection{Reinforcement learning circuit control}
\label{sec:rl_circuit_control}
\begin{figure}[t]
    \centering
    \includegraphics[width=\linewidth]{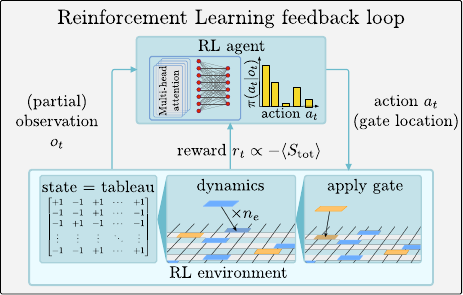}
    \caption{
    \textbf{Schematic Reinforcement Learning (RL) feedback loop.} 
    We train the agent using Proximal Policy Optimization, an actor-critic algorithm where separate transformer-based neural networks model both the actor and critic.
    The input is sampled from $o_t \sim O(o|s_t)$, where $O$ maps the state $s_t$ (Clifford tableau) to a partial observation $o_t$. 
    The agent outputs a policy $\pi(a_t|o_t)$, a probability distribution over locations on the chain. 
    A location $a$ is then sampled and used to place the optimally disentangling two-qubit gate.
    Then, the RL environment evolves by placing $n_{\mathrm{e}}$ random Clifford gates at randomly chosen positions, where $n_{\mathrm{e}}$ is drawn according to $n_{\mathrm{e}} {\sim} (1-p)^{n_\mathrm{e}}$
    [see Fig.~\ref{fig:model}]. 
    The reward $r_t{\propto}\langle \Stot \rangle$ incentivizes the agent to minimize the total entanglement.
    }
    \label{fig:rl_scheme}
\end{figure}

Reinforcement learning (RL) provides a general framework for learning optimal strategies in sequential decision-making tasks~\cite{sutton_2018}.
Such a task is specified through a scalar reward signal $r$, which must be carefully designed to guide the algorithm towards the intended goal.

The RL feedback loop, illustrated in Fig.~\ref{fig:rl_scheme}, has two main components: the RL agent and the RL \textit{environment}~\footnote{The RL environment is conceptually distinct from the notion of environment in open systems in physics.} -- the simulated stochastic stabilizer circuit.
At each turn $\tau$, the RL agent receives a partial observation sampled from $o_t \sim O(o|s_t)$, where $O$ maps states to partial observations by randomly removing rows of the tableau with rate $1-q$ (see Sec.~\ref{sec:partial_info}).
The parameter $q$ controls the amount of information, representing the probability of keeping each stabilizer, i.e., tableau row (see Sec.~\ref{sec:partial_info}).
Based on $o_t$, the RL agent takes an action $a_t$, corresponding to the location of the optimal disentangling gate, which modifies the environment state.
The environment generates a reward signal $r_t$, defined to be proportional to the negative total entanglement:
\begin{equation}
    r_t = - \frac{\langle \Stot \rangle}{\mathcal{N}} \; .
\end{equation}
This reward structure, when maximized, encourages the agent to globally suppress entanglement and actively inhibit its spread across the system.
The environment then undergoes stochastic evolution by placing $n_e$ randomly chosen two-qubit Clifford gates at random locations on the chain, with $n_e$ sampled from the distribution $n_e \sim (1-p)^{n_e}$.

The RL agent's action selection is governed by its policy $\pi_\theta(a_t|o_t)$, a discrete probability distribution over actions $a_t$ conditioned on the current observation $o_t$; it is parameterized by variational parameters $\theta$. 
This formulation naturally generalizes the rule-based strategies discussed above in Eqs.~\eqref{eq:random_policy}, \eqref{eq:greedy_policy}, expressed as probability distributions over the action space.

Learning proceeds in training iterations called \textit{episodes}.
At the start of each episode, the environment is reset to the initial state $\ket{\psi_0}=\ps$ and evolves according to the RL loop in Fig.~\ref{fig:rl_scheme}.
After the episode, the parameters $\theta$ are updated to maximize the expected reward: $\mathbb{E}_{\pi_\theta,\mathcal{P}}[\sum_t r_t]$, where the expectation is taken over the stochastic policy $\pi_\theta$ and the environment's transition probabilities $\mathcal{P}$. 
We use a custom GPU-accelerated implementation of Proximal Policy Optimization (PPO) \cite{schulman2017proximal,brax2021github} (see App.~\ref{app:rl_ppo} for details), a state-of-the-art actor-critic algorithm \cite{konda_1999}, that employs two neural networks: an actor, which defines the strategy $\pi_\theta$, and a critic, which estimates the expected return of actions to guide policy improvements and reduce variance during training.
PPO is particularly well-suited for stochastic, high-variance environments such as our stabilizer circuit dynamics (see Sec.~\ref{sec:model}). 
In our implementation, both the actor and the critic network use separate architectures composed of alternating multi-head self-attention layers followed by shallow feedforward networks and normalization layers.
A more detailed description of the RL implementation details, together with the hyperparameters, is given in the App.~\ref{app:rl}.

Despite the seemingly similar approach of Ref.~\cite{tashev2024}, where they train a RL agent to disentangle, the present work differs substantially.
In their work, the agent acts on a random initial state with the objective of disentangling it. 
By contrast, our setting involves a real-time entanglement-suppression task, where the agent must continuously counteract a stochastic environment that dynamically entangles the system. 
Furthermore, the system sizes considered are different: Ref.~\cite{tashev2024} focuses on small systems of up to five qubits, whereas our study targets genuinely many-body regimes with up to $N = 128$ qubits. \\

The computational cost of the RL algorithm itself scales polynomially with system size $N$, and the agent architecture scales sublinearly (see App.~\ref{app:comp_resources}). 
The dominant bottleneck is the transient time required for the dynamics to reach steady state, which grows as $T_{\mathrm{tr}} \propto N^{\alpha(p)}$, where $\alpha=2$ in the volume-law phase, $\alpha=1$ in the area-law phase, and $\alpha(p)$ exhibits a peak at the transition (see App.~\ref{app:transient_times}). 
This significantly increases the time required for training, especially in the scarce-resource regime $p < p_c^{\mathrm{greedy}}$.
The largest system size we investigate, $N{=}128$, is accessible thanks to a custom-written high-performance implementation of the environment in JAX~\cite{jax2018github}, which enables end-to-end GPU execution and parallelization of thousands of trajectories (see App.~\ref{app:rl_environment}). 
Reaching larger system sizes (e.g., $N=256$) is in principle feasible, though it would entail increased computational wall-clock time. 
Further \textit{ad hoc} improvements to the RL framework and implementation -- such as optimizing data handling -- could help reduce training time.

\subsection{Fully informed entanglement-reducing strategies}
\label{sec:informed_strategies}

\begin{figure*}[t]
    \centering     \includegraphics[width=\textwidth]{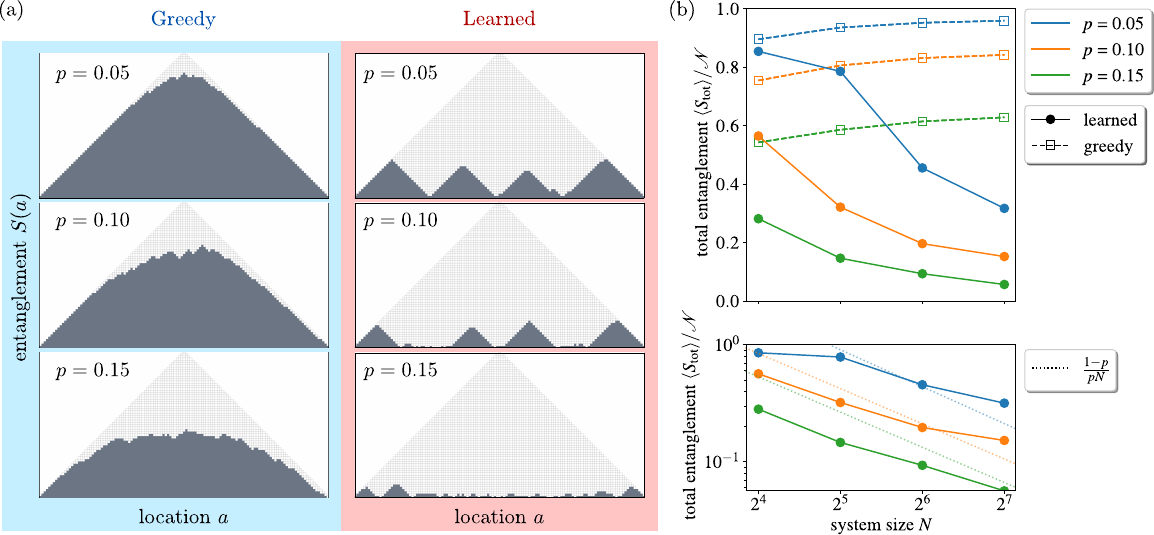}
    \caption{
    \textbf{Greedy vs.~learned active-feedback control strategies.}
    (a) Snapshots of the spatial distribution of entanglement in the steady state for $N=128$ and $p=0.05,0.1,0.15$: greedy (blue, left) and learned (red, right) strategies. 
    Entanglement entropy $S(a)$ at location $a$ on the chain is represented by dark gray blocks (see Sec.~\ref{sec:model}). 
    The greedy strategy focuses on the center of the chain, achieving $\langle \Stot \rangle / \mathcal{N} < 1$ also for $p<\pc^\mathrm{greedy}$. The learned active-feedback-control strategies implement entanglement bottlenecks (``valleys'' between pyramids) to split the system into smaller subsystems, achieving a lower overall entanglement with fewer actions.
    (b) Normalized ensemble-averaged entanglement $\langle \Stot \rangle / \mathcal{N}$ for selected values of the bias parameter $p=(0.05,0.1,0.15)$ as a function of system size $N$, highlighting the qualitative difference between greedy and learned strategies:
    the learned active-feedback-control strategies consistently achieve lower $\langle \Stot \rangle / \mathcal{N}$ that decreases with increasing $N$ across all studied values of $p$, consistent with the absence of a volume-law phase. In contrast, the greedy strategy saturates to finite non-zero values of $\langle \Stot \rangle / \mathcal{N}$, indicating the persistence of volume-law entanglement scaling in the thermodynamic limit.
    The learned strategy data suggest approximate $(pN)^{-1}$ scaling, though deviations occur due to finite-size effects and the inherent variability of learned models [see text]; this scaling is expected to hold only asymptotically.
    }
    \label{fig:rl_fullinfo}    
\end{figure*}

We first investigate the learning of (near-)optimal entanglement-reducing strategies using reinforcement learning (RL) under complete information, where the agent has full access to the quantum state ($o_t = s_t$). 
We aim to uncover beyond-greedy entanglement-reducing strategies that reveal underlying physical insight and novel mechanisms.
Finally, we assess the role of active feedback -- inherent to RL -- in steering the steady state of a quantum system. \\

\textbf{Learned entanglement-reducing strategies.---}%
For each $(p, N)$ pair, we train a separate RL agent as different parameter combinations lead to different optimal strategies that must be learned separately (App.~\ref{app:rl}); we then compute $\Stota$ using the procedure outlined in Sec.~\ref{sec:model}. 
The learned strategies (red circles, Fig.~\ref{fig:combined}) achieve significantly lower steady-state entanglement than the greedy strategy, especially for larger systems ($N=64,128$), demonstrating superior disentangling efficiency. 
As $N \rightarrow \infty$, the data suggest the absence of a critical point, shrinking the volume-law phase to a single point $p=0$ and resulting in a nonequilibrium area-law steady-state for any $p>0$.
This trend is highlighted in Fig.~\ref{fig:rl_fullinfo}(b), showing the normalized total steady-state entanglement for both greedy (squares) and learned (circles) strategies at $p=0.05,0.1,0.15$ and $N=16,32,64,128$.
The learned strategy consistently outperforms the greedy strategy, with the performance gap widening as system size $N$ increases. 
For $p=0.05,0.1$, greedy values plateau at a finite nonzero entanglement value while learned strategies achieve progressively lower entanglement with increasing $N$, suggesting they successfully prevent extensive entanglement growth and spreading in the thermodynamic limit.  

Given the stark performance difference, we investigate the mechanisms behind the high effectiveness of learned strategies by analyzing their steady-state dynamics (see supplementary movies, App.~\ref{app:captions}).
Figure~\ref{fig:rl_fullinfo}(a), red region (right), illustrates snapshots at $p=0.05,0.10,0.15$ and system size $N=128$.
At $p > \pc^\mathrm{greedy}$, no evident control pattern emerges. 
While some pyramidal structures appear near the boundaries (see discussion below), the learned strategy primarily suppresses entanglement uniformly, similar to the greedy strategy but without its spatial asymmetry.
In this regime, the greedy strategy is nearly optimal, achieving area-law scaling; the learned strategy mimics it, with minor refinements.

Clear control patterns emerge in the `scarce-resource' low-$p$ regime, $p < \pc^\mathrm{greedy}$, where acting greedily is no longer optimal.
In this setting, the learned strategy is qualitatively different: it targets specific bonds -- or clusters of bonds -- that are approximately equidistant in space, and actively suppresses entanglement at these locations. 
Maintaining near-zero entanglement on these selected bonds creates entanglement bottlenecks that prevent its spread across the system, effectively splitting the chain into subsystems of maximal entanglement.
This leads to the occurrence of stable pyramidal structures in the spatial entanglement profile of the long-time steady state, as seen in Fig.~\ref{fig:rl_fullinfo}(a) (right, red region).
This approach is remarkably efficient, limiting entanglement growth with a relatively low bias $p$, even as the environment continuously attempts to uniformly entangle the system. 

The comparison between greedy and learned strategies underscores the importance of active feedback control. The framework is not only effective in the specific context of learning entanglement-reducing strategies, but it also points to a broader conclusion: active feedback, implemented through reinforcement learning, can serve as a powerful and efficient tool for controlling quantum many-body systems. We believe this insight extends beyond our current setting and may hold significant potential even in more general scenarios involving interactions with quantum many-body systems.\\

\textbf{Importance of real-time active feedback.---}%
These learned strategies appear simple, yet their effectiveness proves difficult to replicate with simple, human-designed rules.
To illustrate this, we consider two such strategies (see App.~\ref{app:human_designed_strategies} for details).
The first is a ``pyramid strategy'' that aims to remove entanglement at $n_\mathrm{B}$ equidistant bonds and acts on these plus neighboring bonds ($3$ bonds per bottleneck) whenever entanglement can be reduced.
The second is a more sophisticated ``smart strategy'', suggested by an anonymous referee: rather than targeting fixed bottleneck locations, it aims to maximize the inactive regions, i.e., regions of maximal steepness in the entanglement profile where entanglement can be neither created nor removed, while creating as many bottlenecks as possible.

Our numerical simulations (see App.~\ref{app:human_designed_strategies}) show that the pyramid strategy initially maintains $\Stot/\mathcal{N} < 1$ (higher than learned strategies, yet not maximal), but it proves unstable: bottlenecks are lost and the total entanglement keeps growing over time.
The smart strategy also fails to sustain the pyramidal structure in the entanglement profile, and the system approaches a near-maximally entangled state.
Hence, these simple rules are not sufficient to reproduce the area-law steady states of the learned strategies.
These results suggest that reproducing the learned strategies is not trivial, and that counteracting entanglement growth benefits from active feedback acting on a sufficiently long horizon\footnote{
  Greedy, human-designed, and learned strategies are all feedback strategies in the proper sense: each observes the current state and acts on it online.
  The greedy strategy is myopic: it only considers the entanglement reduction at the current turn.
  The learned agent, trained with a high discount factor $\gamma = 0.99$, is instead optimized for the long-term return, and can forgo immediate reduction in order to establish and maintain bottlenecks whose benefit is realized many turns later.
  The smart strategy also pursues a long-term goal, but one designed by hand and formulated on the entanglement profile alone.
}.
Our results do not, however, exclude that more sophisticated rule-based strategies could achieve a comparable performance. \\

\textbf{Area-law entanglement scaling.---}%
Let us now argue that, under the learned strategy, the results are consistent with area-law scaling of entanglement in the steady state. 
First, the pyramid structures observed in the learned strategy exhibit self-similar scaling behavior: when doubling the system size $N \to 2N$, the optimal strategy accommodates twice as many pyramids, as the action density of the stochastic evolution also halves. 
Evidence of self-similarity appears in our numerical data: as the system size doubles from $N=64$ to $N=128$ at $p=0.05$, the entanglement profile also doubles in structure, increasing from $2$ to $4$ pyramid structures (see supplementary videos, App.~\ref{app:captions}).
As a result, the total entanglement doubles: $S_{\text{tot}}^{(2N)} = 2 S_{\text{tot}}^{(N)}$, giving $S_{\text{tot}} \sim N$. However, the normalization in Eq.~\eqref{eq:normalization} grows quadratically, $\mathcal{N} \sim N^2$. Hence, the density $S_{\text{tot}}/\mathcal{N} {\sim} N^{-1}$, confirming area-law scaling.

To better quantify the learned behavior and its scaling laws, we introduce a simplified model of the dynamics under the learned strategy; while disregarding some underlying physics, it captures essential features and provides useful intuition of the phenomenon.
Consider a one-dimensional chain of $N-1$ bonds ($N$ sites). At each time step, the stochastic evolution places a random gate with probability $1-p$, whereas the strategy places a disentangling gate with probability $p$. 
The latter prevents entanglement spreading by implementing $n_\mathrm{B}$ entanglement bottlenecks.
At a fixed time step, the probability of reducing/increasing the total entanglement $\Stot$ consists of the product of the probability for the strategy/stochastic dynamics to place a gate and the corresponding available bond density.
While the strategy only acts on the $n_\mathrm{B}$ bottleneck bonds, the stochastic dynamics places gates on any one of the $N-1$ bonds.
For these bottlenecks to be stable in the steady state requires matching the probability of reducing/increasing the total entanglement.
Importantly, in the steady state, the pyramid structures lead to the emergent constraint that entanglement can only be created at the bottlenecks, since the state is locally maximally entangled everywhere else, leading to an asymmetry in the available bond densities. Thus, the steady state condition reads as
\begin{equation}
    p \times \frac{\cancel{n_\mathrm{B}}}{\cancel{n_\mathrm{B}}} = (1-p)\times \frac{n_\mathrm{B}}{N-1} .
    \label{eq:prob_relation}
\end{equation}
Next, by counting the number of pyramids and their volume, we find that the normalized entanglement scales as ${\Stot}/{\mathcal{N}} \simeq {1}/{(n_\mathrm{B} + 1)}$.
Using Eq.~\eqref{eq:prob_relation} yields: 
\begin{equation}
    \frac{\Stot}{\mathcal{N}} = \frac{1}{\frac{p}{1-p}(N-1) + 1}\sim \frac{1-p}{pN}, 
    \label{eq:s_tot_scaling}
\end{equation}
in the thermodynamic limit $N \rightarrow \infty$.
This shows that the total entanglement vanishes as $N\rightarrow\infty$ for any 
fixed $p>0$. 

We superimpose this scaling onto the learned strategy plot in Fig.~\ref{fig:rl_fullinfo}(b), bottom plot.
Some deviations are observed, particularly for $p = 0.05$ and $0.1$, as finite-size effects play a significant role: fitting the system with a fixed number of pyramids may be suboptimal, and the RL framework converges to a strategy that also eliminates some pyramids (e.g., Fig.~\ref{fig:rl_fullinfo}(b), $p=0.10$), which alters the results. 
Moreover, learning strategies that fit the system with the correct number of equidistant pyramids involves converging to a specific local minimum in a complex loss landscape -- a fine-tuning task we do not pursue here.

Additionally, the simplified model introduced above neglects certain effects from the full dynamics, including correlations between neighboring locations and higher-order processes such as simultaneous entanglement creation at adjacent sites, which can prevent disentangling. These omitted physical processes contribute to the observed deviations from the theoretical prediction.
Nevertheless, the simplified model captures the essential scaling behavior and overall trends; such scaling is expected to hold only asymptotically, providing valuable insight into the thermodynamic limit of the learned strategy. \\ 

\subsection{Partially informed strategies}
\label{sec:partial_info}
\begin{figure*}[t]
    \centering
    \includegraphics[width=\textwidth]{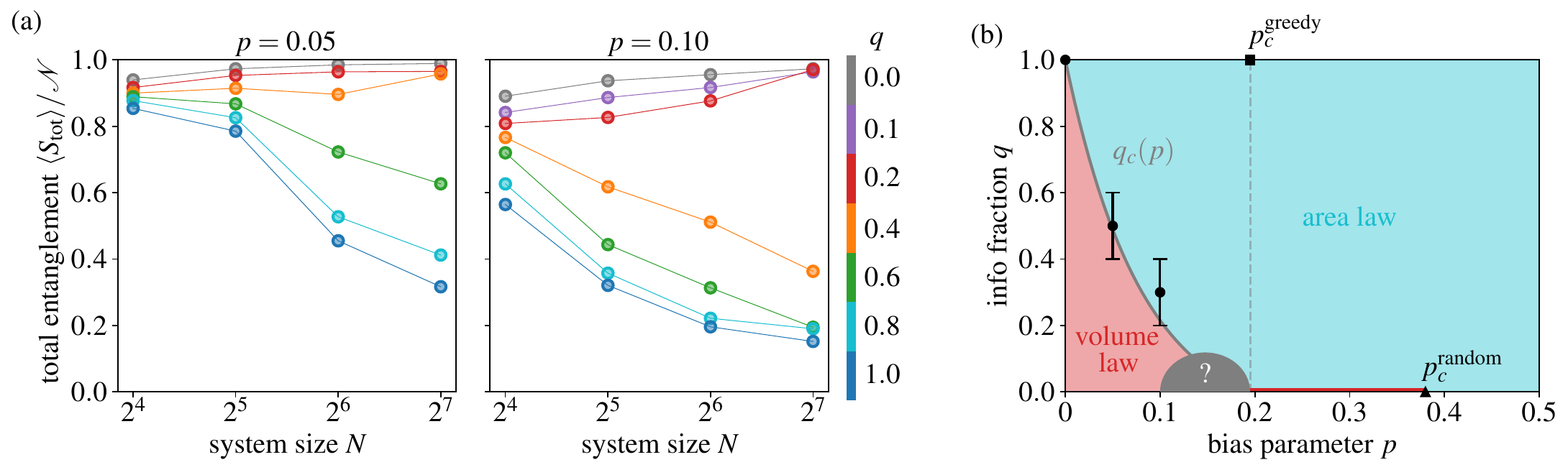}
    \caption{
    \textbf{Partially informed learned strategies.}
    (a) Normalized ensemble average $\Stota$ for learned policies with partial state information parametrized by $q$ (colorbar).
    Each point represents a single trained model.
    For increasing system size $N$, the bifurcation reveals a critical value $q_c(p)$ below which efficient entanglement-reducing strategies cannot be implemented, and the long-time steady states feature volume-law entanglement scaling.
    (b) Phase diagram of the information fraction $q$ vs.~the bias parameter $p$.
    Learned strategies achieve area-law scaling throughout most of the phase diagram, with a transition to the volume-law phase.
    Circle data points with error bars from data in panel~(a); error bars reflect the discretization of $q$.
    The data point at $(p,q)=(0,1)$ corresponds to maximally entangled states at zero bias $p$.
    The curve $q_c(p)$ is a schematic guide-to-the-eye highlighting the phase boundary.
    The square (triangle) marks the critical point of the greedy (random) strategy.
    The red line at $q=0$ indicates a volume-law phase at zero information up to $p_c^\mathrm{random}$, whereas area-law scaling is recovered for any $q>0$ when $p \geq p_c^\mathrm{greedy}$.
    The grey area indicates where finite-size effects are too strong to draw quantitative conclusions.
    }
    \label{fig:rl_p}
\end{figure*}
Learned strategies leveraging complete quantum state information via active feedback have proven to be both efficient and effective. 
This raises a natural question: is access to the complete state information truly necessary to implement such strategies?
This is particularly relevant in realistic experimental settings, where the complete quantum state is inaccessible and only partial state information is available~\cite{cramer_2010,toth_2010,lanyon_2017,rocchetto_2019}.

Let us now investigate whether learned strategies can prevent extensive entanglement in the steady state with partial quantum state information.
We restrict only the information available to the entanglement-reducing strategy, while optimally disentangling gates are still determined from the local state [cf. Sec.~\ref{sec:model}].
We model this by training RL agents on partial observations $o_t$, sampled according to $o_t \sim O(o|s_t)$; $O$ maps the full state tableau $s_t$ to partial observations, sampling each stabilizer in the tableau independently with probability $q \in [0,1]$. 
Operationally, this partial access to stabilizers corresponds to working with an effective mixed state $\rho$ with reduced purity $\Tr \rho^2 = 2^{-N(1-q)}$. 
As in the previous section, we train a separate RL agent for every triple $(p, N, q)$.
Rather than training each agent \textit{ab initio}, we employ curriculum learning, which significantly accelerates convergence: starting from a fully informed model at $q{=}1$, we progressively reduce the information rate $q$, using the trained agent at each step as an initialization for the training at the next lower value of $q$.
We then compute $\Stota$ using the procedure outlined in Sec.~\ref{sec:model}. 
While agents can be trained independently for each value of $q$, the curriculum approach reduces both training time, the need for extensive hyperparameter tuning, and reseeding stability.

Consider the scarce-resource regime $p < p_c^\mathrm{greedy}$ where fully informed learned strategies exhibit pyramidal structures in the spatial entanglement profile ($p = 0.05,0.10$). 
Figure~\ref{fig:rl_p}(a) shows the results of training RL agents with varying levels of partial information $q$ (see colorbar). 
The data displays a clear bifurcation in the scaling of the total entanglement $\Stota/\mathcal{N}$ with system size $N$, indicating the presence of a critical information threshold $q_c(p)$. 
Above the threshold, $q > q_c(p)$, the accessible information is sufficient to prevent entanglement spreading, and long-time steady states feature area-law entanglement scaling. Below the threshold, $q < q_c(p)$, insufficient information causes the entanglement-reducing strategy to fail, reverting to volume-law entanglement scaling.
Notably, increasing the bias $p$ reduces the amount of information required to achieve effective control, reflecting a clear trade-off: a weaker bias towards the entanglement-reducing strategy requires more information to steer the steady state efficiently.

In the regime $p \geq \pc^\mathrm{greedy}$, learned strategies with partial information exhibit area-law scaling even for the smallest $q{=}0.1$ value investigated.
This leads us to conjecture that area-law behavior may persist in the limit $q \to 0$. 
As system size increases, the gate density decreases, and entanglement predominantly comes from nearest-neighbor Bell-like pairs~\footnote{A Bell-like pair is a two-qubit state that is maximally entangled and locally equivalent (up to single-qubit Clifford operation) to a Bell pair.} -- since only nearest-neighbor gates are applied. 
In the vanishing-information limit ($q \to 0$), each observation consists of a single stabilizer providing localized information about one Bell-like pair. 
This information can always be exploited to eliminate the pair before it spreads entanglement over larger regions, amounting to a greedy-like strategy.
Numerical evidence supports this picture. We empirically extract the conditional probability of targeting a Bell-like pair given that the observation contains relevant information: for learned strategies at $N{=}128,p{=}0.2$, this probability amounts to $76\%$ at $q{=}0.2$ and $71\%$ at $q{=}0.1$. We find that both strategies achieve area-law scaling.
While additional mechanisms may contribute to the overall behavior, this consistently high targeting efficiency supports the conjecture that area-law entanglement scaling is achievable for any $q>0$ when $p \geq \pc^\mathrm{greedy}$ in the thermodynamic limit.

Our findings are summarized in the phase diagram of bias parameter $p$ against information fraction $q$, shown in Fig.~\ref{fig:rl_p}(b).
The data points with error bars correspond to those in Fig.~\ref{fig:rl_p}(a), while the square and triangle mark the critical points of the greedy and random strategies, respectively.
A guide-to-the-eye curve highlights the trend of the critical information threshold $q_c(p)$.
The red line at $q=0$ indicates that, with strictly zero information, the learned strategy stabilizes volume-law scaling up to $p_c^\mathrm{random}$ and area-law scaling for any $q>0$ when $p \geq p_c^\mathrm{greedy}$.
The grey region marks where finite-size effects are too strong to draw quantitative conclusions (see App.~\ref{app:vanishing_info} for further discussion).

\section{Discussion and Outlook}
\label{sec:discussion_outlook}

In this work, we investigate the role of information through active feedback to control entanglement in nonequilibrium steady states of stabilizer circuits. 
By leveraging reinforcement learning (RL), we identify strategies that drive the system out of equilibrium and achieve area-law entanglement, featuring nontrivial steady-state distributions otherwise impossible in equilibrium. 
These include spatially localized entanglement profiles and bottlenecks, which differ qualitatively from random and greedy strategies.
Active feedback, which accounts for the long-term consequences of each action, is an important component of the learned strategies; the simple human-designed strategies we tested could not reproduce their performance and fail to sustain out-of-equilibrium steady states.
In particular, our results about learned strategies are consistent with eliminating the volume-law phase entirely: even an infinitesimal bias $p{>}0$ is sufficient to drive the system into an area-law steady state.
Unlike previous studies employing RL within the mean-field regime or for coarse-grained control~\cite{bukov_2018,wauters_2020,guo_2021,metz_2023}, our framework directly controls the dynamics of an extensive number of individual degrees of freedom.
These results advance both the conceptual understanding and technical realization of information-driven quantum many-body control.

A key insight of our work is that RL agents learn and implement entanglement-reducing strategies even under partial information, provided the information fraction $q$ exceeds a threshold $q_{\mathrm{c}}(p)$ which separates two qualitatively distinct regimes: $q{>}q_c(p)$, where entanglement growth is actively inhibited; and $q{<}q_c(p)$, where control fails due to insufficient information and the volume-law phase reemerges. 
This threshold likely reflects a transition in the space of accessible control protocols driven by the amount of available information.

Our findings contribute to a growing body of work exploring learning-driven control in quantum many-body systems~\cite{bukov_2018,fosel_2018,wauters_2020,guo_2021,metz_2023,tashev2024,olle_2024}.
The emergence of an information threshold $q_\mathrm{c}(p)$ in our system suggests a transition reminiscent of learning-induced phenomena -- phase transitions in the ability of learning to extract or manipulate global properties from local data \cite{putz_2025}.
Whether this learning transition is universal or model-specific remains an open question.
Future research will characterize critical properties, such as critical exponents and dynamical scaling behavior near $\pc^\mathrm{greedy}$, for both greedy and learned strategies, and at the critical information threshold $q_{\mathrm{c}}(p)$. 

In the broader context of strategy discovery, our study uncovers a universal principle: when the game is not even, e.g., the agent plays once every twenty moves on average, or competes with limited resources against a much more resourceful opponent, the learned strategy seeks a state with emergent constraints -- here pyramids separated by entanglement bottlenecks. 
In practice, this effectively reduces the opponent's capabilities: the environment can place an entangling gate anywhere, but the entanglement itself increases only if the gate is placed on (or close to) a bottleneck bond, since the surrounding subsystems are already maximally entangled.
The learned strategy, however, leverages a different resource: it uses information to first induce and then exploit these constraints by identifying effective control degrees of freedom (the bottlenecks) in the underlying dynamics.
In the strongly biased regime, a non-vanishing minimal amount of information is required to make up for the unevenness of the game.
This reveals a generic principle of learning optimal strategies that applies beyond the scope of the present work, and suggests several avenues for future studies. \\

Immediate research directions include an extension to higher-dimensional systems and the exploration of how the pyramidal entanglement structures observed in learned strategies generalize to more complex settings.
We therefore distinguish which of our findings are specific to the stabilizer structure and which we expect to extend more broadly.
Features that are specific to Clifford circuits or the tableau representation are:
the quantization of the entanglement, $S(a)=n(a)\ln 2$, underlying the block representation of the entanglement profile and the counting argument leading to Eq.~\eqref{eq:s_tot_scaling}; the existence of an exactly optimal two-qubit disentangling gate, obtained by lookup rather than by continuous optimization (App.~\ref{app:disentangling_cliffords}); the Bell-pair targeting mechanism identified in the small-$q$ limit [Sec.~\ref{sec:partial_info}]; the numerical values of $p_c^{\rm random}$ and $p_c^{\rm greedy}$ and the operational definition of the information fraction $q$ as stabilizer removal.
These features would need to be revised before application on other circuit classes.
By contrast, the nonequilibrium character of the feedback-induced steady state
(App.~\ref{app:nonequilibrium}) follows from information-based biasing of transitions and makes no reference to the gate set. The same holds for the strategic principle discussed above: an agent acting with scarce resources induces emergent constraints that reduce the effective action space of the stochastic dynamics.
Consequently, the trade-off between control frequency and available information is expected to be general, even though its quantitative form, such as the boundary $q_c(p)$, is model dependent.

A natural question concerns Haar-random circuits, 
where previous work has shown the absence of a transition in the thermodynamic limit and maximally entangled steady states for all $p{<}1$~\cite{yepes_2024}.
We expect the coarse mechanism -- partitioning the chain by suppressing entanglement at a sparse set of locations -- to remain a sensible strategy there, since it exploits the locality of the gates rather than any property of the Clifford group.
The details, however, may change appreciably.
In Haar-random circuits, entanglement is continuous: a disentangler can reduce it, but generically cannot pin the entanglement across a bond to zero. 
Bottlenecks would then be local minima with nonzero entanglement. 
Whether active feedback can nonetheless stabilize area-law steady states in this setting is an open question that requires a dedicated study.
It is also numerically more demanding, as simulating generic quantum circuits becomes exponentially harder with system size, limiting the accessible system sizes, while both the gate optimization and the location selection become substantially more difficult problems.

Translating such strategies into actual experiments would be desirable.
In such an experiment, the stabilizer circuit is run on hardware while tracking the tableau classically, with $q$ interpreted as the fraction of stabilizers accessible per time step -- since stabilizers commute, they can all be read out simultaneously.
Another relevant experimental setting involves implementing Haar-random circuits.
Here, information is acquired through measurements -- ancilla-based or weak measurements -- whose outcomes serve as the agent's observation, and the action space comprises gate applications and/or measurement choices. Moreover, the objective itself need not be entanglement suppression.
Our findings suggest that relying on single-shot measurements -- rather than full quantum state tomography -- may suffice, since learned strategies succeed even with partial state knowledge.
At the same time, implementing RL-based feedback on quantum hardware remains challenging, requiring either fast access to classical measurement outcomes, leveraging hybrid classical--quantum approaches, or direct use of quantum data -- e.g., via quantum agents or hybrid architectures~\cite{reuer_2023,nagy_2024}.

While our focus has been on reducing entanglement, the framework developed here -- classical feedback-based control using partial quantum state information -- has much wider applicability.
The same principles could be adapted to stabilize nonequilibrium steady states with novel critical properties or quantum many-body phases of matter, such as topological phases~\cite{wen_2017}, symmetry-protected topological phases~\cite{senthil_2015}, or many-body localized steady states~\cite{nandkishore_2015}. 
Moreover, active feedback could be exploited for state preparation~\cite{zhou_2019}, quantum memory~\cite{terhal_2015} protocols, and quantum error correction~\cite{wang_2024}, where adaptive strategies could help detect and mitigate errors under limited information. 
In this sense, our work provides a concrete and accessible foundation for active, information-driven control of quantum dynamics in far-from-equilibrium systems with many interacting degrees of freedom.
Taken together, these directions point towards a broader paradigm of information-driven quantum control, where learning and feedback play a central role in navigating complex many-body dynamics.

\section*{Acknowledgments}
We acknowledge fruitful discussions with Jiangtian Yao, Gianluca Teza, and Hansveer Singh.
We are grateful to an anonymous Referee for suggesting one of the human-designed strategies and for providing the corresponding implementation.
The authors used Claude Opus 4.8/5 during the preparation of this manuscript to improve readability. 
The authors assume full responsibility for the scientific content, accuracy, and integrity of this paper.  
Funded by the Deutsche Forschungsgemeinschaft (DFG, German Research
Foundation) -- 544919793, and the European Union (ERC, QuSimCtrl, 101113633). Views and opinions expressed are, however, those of the authors only and do not necessarily reflect those of the European Union or the European Research Council Executive Agency. Neither the European Union nor the granting authority can be held responsible for them.
MS was supported through the Helmholtz Initiative and Networking Fund, Grant No.~VH-NG-1711.
Numerical simulations were performed on the MPCDF HPC cluster. 

\textit{Data and Code Availability.---}%
The data, code, and trained models associated with this manuscript version are available under DOI:\href{https://doi.org/10.5281/zenodo.23806496}{10.5281/zenodo.23806496}~\cite{qtetrix_repo}.
Automatic access is restricted; access credentials will be provided to human researchers upon reasonable request via Zenodo for academic and verification purposes.

\clearpage
\begin{appendix}
\resumetoc
\tableofcontents

\section{Description of supplementary videos}
\label{app:captions}
\begin{figure}[t]
    \includegraphics[width=\linewidth]{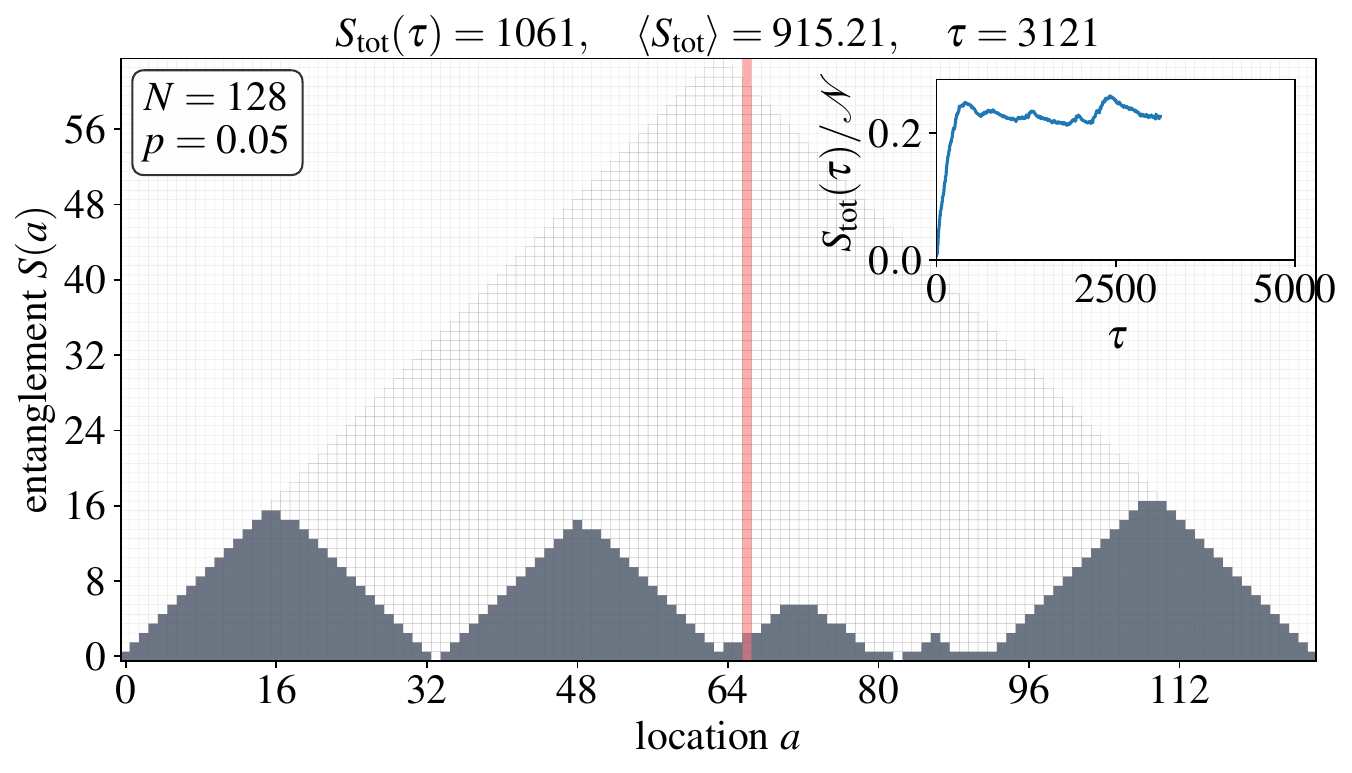}
    \caption{
    \textbf{Representative snapshot from supplementary video \textit{learned\_128\_050.mp4}.}
    The frame shows the entanglement profile at turn $\tau$ during the evolution under the learned strategy.  
    Highlighted in red is the location of the disentangling gate.  
    Displayed at the top are the current total entanglement $\Stot(\tau)$, its time average $\Stota$, and the turn number $\tau$.  
    Darker lines in the grid mark the maximum possible entanglement; dark gray rectangles show the spatial profile of entanglement in the current state. The inset in the top-right corner shows the normalized entanglement $\Stot(\tau)/\mathcal{N}$ over time.
    }
    \label{fig:example_snapshot}
\end{figure}

The paper is accompanied by twelve supplementary videos illustrating the dynamics of the entanglement profile [see Fig.\ref{fig:model}(b)] for both greedy and learned strategies, starting from the product state $\ps$ (see ancillary files on arXiv).
Each frame of the video corresponds to a single turn $\tau$ of the strategy.
A representative example of such a frame is shown in Fig.\ref{fig:example_snapshot}.
Highlighted in red is the location of the next disentangling gate. 
At the top of the screen, we display the instantaneous total entanglement $\Stot(\tau)$, its running time average $\Stota$, and the current turn number $\tau$. 
Darker lines in the grid mark the maximum possible entanglement in the system. 
The top right corner shows a live plot of the normalized total entanglement  $\Stot(\tau)/\mathcal{N}$.
The filenames follow the convention \textit{nameOfStrategy\_$N$\_$P$.mp4}, where $P = 1000\,p$ is written with three digits. For example, the video of the learned strategy for $N=128$ and $p=0.05$ is called \textit{learned\_128\_050.mp4}.

Caption for each video:
\begin{itemize}
    \item \textit{greedy\_128\_050.mp4} -- The video shows fast entanglement spreading and the inability of the greedy agent to efficiently counteract the stochastic evolution, leading to high values of total entanglement.
    There is a transient period ($\tau \lesssim 2000$) where entanglement grows above $S_{\text{tot}} > 3800$. 
    The greedy strategy prevents maximal entanglement $\Stot/\mathcal{N} = 1$ by removing the top part of the pyramid.
    Similar behavior is observed in \textit{greedy\_64\_050.mp4} with a shorter transient period ($\tau \lesssim 600$) and lower peak entanglement values ($S_{\text{tot}} > 900$).
    
    \item \textit{greedy\_128\_100.mp4} -- The video is similar to the previous one, with the greedy strategy failing to efficiently inhibit entanglement spreading: the transient phase is significantly longer ($\tau \lesssim 5000$) with entanglement growing above $S_{\text{tot}} > 3200$. 
    In the steady state, a larger fraction of the top part is removed compared to the $p=0.05$ case.
    Similar behavior is observed in \textit{greedy\_64\_100.mp4} with a shorter transient period ($\tau \lesssim 1700$) and lower peak entanglement values ($S_{\text{tot}} > 800$).

    \item \textit{greedy\_128\_200.mp4} -- As $p=0.2$, frequent disentangling operations enable the greedy strategy to prevent entanglement spreading, effectively maintaining $S_{\text{tot}} < 300$ throughout most of the dynamics. Similar behavior is observed in \textit{greedy\_64\_200.mp4}.

    \item \textit{learned\_128\_050.mp4} -- This video shows the formation of pyramid-shaped entanglement profiles, reflecting the entanglement bottleneck mechanism of the learned strategy. 
    After a short transient period ($\tau \lesssim 600$), entanglement grows above $S_{\text{tot}} > 900$, yet remains significantly lower than the greedy strategy. The strategy maintains a four-pyramid arrangement in the steady state, focusing on disentangling the valleys between pyramids to achieve low total entanglement. In \textit{learned\_64\_050.mp4}, only two pyramid-shaped entanglement profiles form after a longer transient period ($\tau \lesssim 700$), with the system exhibiting lower steady-state entanglement ($S_{\text{tot}} \approx 430$).
    
    \item \textit{learned\_128\_100.mp4} -- This video shows pyramid formation with $p=0.1$: compared to $p=0.05$, the pyramids are smaller and separated by entire subregions near zero entanglement, achieving steady-state entanglement $S_{\text{tot}} < 600$. 
    Similarly in \textit{learned\_64\_100.mp4}, achieving steady-state entanglement ($S_{\text{tot}} < 300$).
    
    \item \textit{learned\_128\_150.mp4} -- This video shows the learned strategy resembling a greedy-like approach since at $p=0.15$ the greedy approach is nearly optimal. However, a couple of pyramids form at the chain boundaries. The learned strategy achieves lower average entanglement than the greedy strategy (see Fig.~\ref{fig:rl_fullinfo}). 
    In \textit{learned\_64\_150.mp4}, the learned strategy also resembles a greedy-like approach, but distinct pyramids are less visible due to the smaller system size and high bias rate, while still outperforming the greedy strategy.
    
\end{itemize}

\section{Nonequilibrium long-time steady states induced by feedback control}
\label{app:nonequilibrium}

In this section, we discuss the nonequilibrium properties of the steady state reached by the circuit dynamics under feedback control.
A steady state occurs when the system's macroscopic properties become time-independent, while microscopic dynamics continue. 
In our case, the steady state is characterized by time-independent values of the ensemble average $\Stota$.
However, this does not imply equilibrium -- the steady state can be maintained by continuous energy and entropy flows, rendering it out-of-equilibrium.

The system described in Sec.~\ref{sec:model} constitutes a feedback control system.
At each turn, information about the quantum state $s$ feeds back to the controlling agent, informing both the strategy $\pi(a|s)$ selecting the location $a$ and the mapping selecting the optimally disentangling gate at the chosen location $a$. 
Such feedback-controlled systems are inherently out of equilibrium and reach nonequilibrium steady states.
The nonequilibrium nature manifests through violations of detailed balance and non-constant thermodynamic quantities, such as non-zero entropy production.
To see that, we consider the composite system comprising the physical quantum circuit, the controlling agent, and the environment. The literature on thermodynamics of feedback-controlled systems is well-established and provides the framework for this analysis \cite{cao_2009,koski_2014,sagawa_2012,prech_2024}.
While the physical system relaxes to a steady state with time-independent average entanglement and a constant probability distribution over the state space, the feedback loop -- consisting of information flowing from circuit to agent and the subsequent consumption of this information -- continuously generates entropy.
Crucially, the agent possesses internal degrees of freedom that consume energy to operate and exchange information with the system.
As for the Maxwell demon problem, the agent must erase its memory after each operational cycle to enable the entire system to operate cyclically -- an irreversible process that generates entropy.
Any agent using information to steer a system must satisfy Landauer's principle; erasing each bit of information costs at least $k_\mathrm{B}T\ln2$ units of dissipated energy, providing a fundamental thermodynamic cost for information-based control.

\input{tikz_diagrams/bipartite_graph}
Moreover, feedback-controlled systems break detailed balance. 
Detailed balance states that, at equilibrium, the local probability flux between any two states must be equal in both directions: $P(i \rightarrow j) \times p_i = P(j \rightarrow i) \times p_j$, where $P(i \rightarrow j)$ is the transition probability and $p_i$ is the steady-state probability of state $i$. This condition is broken whenever information-based control biases state transitions.
While detailed balance violations imply nonequilibrium, directly observing them from data is hard. 
For larger state spaces, where the number of possible configurations grows exponentially with system size, precise estimation of transition probabilities and steady-state distributions requires extensive data collection.
Interestingly, detailed balance may hold for the physical system alone; this requires that the controller engineer the system to mimic thermal equilibrium statistics, even though the underlying process maintaining this state is fundamentally nonequilibrium. 
The result is a NESS that appears thermal when viewed from the perspective of the controlled system only.

To analyze detailed balance violations, we follow the composite system analysis framework presented in \cite{horowitz_2014}.
For simplicity, we start by considering a simplified system: a physical system with only two states $y=0,1$, where $\Stot(0) < \Stot(1)$, and the agent as a two-state system representing the states where the agent has no knowledge ($x=0$) and has knowledge ($x=1$) of the system state, respectively.
This binary reduction captures the essential physics of feedback control -- the agent's ability to bias transitions based on information -- while remaining analytically tractable. 
The key insight is that detailed balance violations arise from the asymmetric transition probabilities when the agent possesses information, regardless of the system's complexity.

The composite system has four states $(x,y)=(0,0),(0,1),(1,0),(1,1)$, as depicted in Fig.~\ref{fig:bipartite_graph}. 
The stochastic evolution of the circuit yields transitions $(0,0) \leftrightarrow (0,1)$. 
However, when the agent knows the state ($x=1$), the transition becomes unidirectional by construction -- entanglement cannot increase because transitions are biased through agent engineering: $P\big( (1,1) \rightarrow (1,0)\big)=1 \neq P\big((1,0) \rightarrow (1,1)\big)=0$.
Then, the agent loses knowledge of the state through memory erasure: $(1,0) \rightarrow (0,0)$.
This operational cycle creates probability currents that violate detailed balance, proving that the system is out of equilibrium. 
The generalization to more than two states is straightforward, with the agent always biasing transitions from higher to lower entanglement.

In the model analyzed in this work (see Sec.~\ref{sec:model}), even when the strategy $\pi$ does not receive state information as input, the system remains nonequilibrium since the gate selection (mapping two-qubit states to disentangling gates) retains access to a subset of the state information. Thus, the evolution dynamics of the circuit state depend on the circuit state itself in a nonlinear way (via the local density matrix used to determine the next disentangling gate); this nonlinearity gives rise to a nonequilibrium steady state similar to Lindblad dynamics.
Generally, any information usage by any component of the controller contributes to the nonequilibrium character of the overall system.
However, for the random strategy in the regime $p<\pc$, the steady state approaches the maximally entangled state, where entanglement can only be removed at locations in the central region of the chain.
Since the probability of randomly sampling this region scales as $\sim 1/N$, it vanishes in the thermodynamic limit, effectively yielding a system at equilibrium.

Equilibrium conditions are too restrictive, and it is the nonequilibrium nature of feedback-controlled systems that enables the modification of the critical properties in the steady state. 
Without the information-driven bias that breaks detailed balance, the system would be constrained to its equilibrium configuration, making targeted entanglement control impossible. 
The nonequilibrium steady state provides the necessary framework for achieving the controllable entanglement dynamics observed in our feedback protocols.

\section{Steady-state sampling of $\langle S_\mathrm{tot} \rangle$}
\label{app:sampling_steady_state}
This section describes the sampling methodology used to estimate the long-time steady-state average $\langle S_\mathrm{tot} \rangle$.
Starting from the product state $\ket{\psi_0} := \ket{0}^{\otimes N}$, we simulate an ensemble of $M$ independent circuits.
At each turn $\tau$, we compute $S_\mathrm{tot}$ -- where a ``turn'' is the interval between consecutive applications of disentangling gates, i.e., one step of the entanglement-reducing strategy. \\

\textbf{Random and greedy strategies}. 
The ensemble size $M$ is chosen based on the system size $N$: larger circuits are more self-averaging and therefore require fewer independent realizations.
The values used are listed in Table~\ref{tab:trajectory_counts}.

\begin{table}[h]
    \centering
    \renewcommand{\arraystretch}{1.3}
    \setlength{\tabcolsep}{8pt}
    \begin{tabular}{l|ccccc}
        \multicolumn{6}{c}{\textit{Random}} \\
        \midrule
        $N$ & $16$ & $32$ & $64$ & $128$ & $256$ \\[2pt]
        $M$ & $1024$ & $1024$ & $512$ & $256$ & $128$ \\[8pt]
        \multicolumn{6}{c}{\textit{Greedy}} \\
        \midrule
        $N$ & $16$ & $32$ & $64$ & $128$ & $256$ \\[2pt]
        $M$ & $4096$ & $2048$ & $1024$ & $512$ & $256$ \\
    \end{tabular}
    \caption{Number of independent trajectories $M$ used to sample the steady state for the random and greedy strategies at different system sizes $N$.}
    \label{tab:trajectory_counts}
\end{table}

The steady-state average $\langle S_{\mathrm{tot}}\rangle$ is estimated by averaging over all $M$ realizations and over the final $10\%$ of each trajectory, provided the stationarity diagnostic described in App.~\ref{app:stationarity_check} is satisfied over that window. \\

\textbf{Learned strategies}. 
Each trained strategy is distinct and may exhibit different transient timescales and fluctuations; we therefore adopt a more conservative approach.
The ensemble size is fixed at $M = 2048$ across all system sizes;
we assume a worst-case transient time scaling with $N^3$ (see App.~\ref{app:transient_times}), and allow each circuit to equilibrate for $n_\text{th} \times N^3$ turns $\tau$ before sampling $S_\mathrm{tot}$.
We then save $10$ samples separated by $10 \times N$ turns to suppress temporal autocorrelations.
The thermalization prefactor $n_\text{th}$ is set as:
\begin{equation}
    n_\text{th} = \begin{cases}
        0.1 & N \in \{16, 32, 64\}, \\
        0.05 & N = 128.
    \end{cases}
\end{equation}
In practice, we found converged RL strategies equilibrate well within these bounds; the conservative transient time is used as a safeguard against slow drift toward the maximally entangled state.

\subsection{Stationarity check}
\label{app:stationarity_check}
To verify that the dynamics has reached steady state, we apply a Geweke-style diagnostic \cite{geweke_1992}.
Each trajectory consists of $(S_\mathrm{tot}, t)$ values, where $t$ is the cumulative gate count.
We partition each trajectory into three non-overlapping segments of the total simulation time $t_\mathrm{max}$:
\begin{itemize}
    \item a \textit{tail} segment spanning the final fraction $f_\text{tail}$ of $t_\text{max}$, over which we estimate $\langle S_\mathrm{tot} \rangle$;
    \item a \textit{reference} segment of width $f_\text{ref}$ ending at $t_\text{max}(1 - \delta)$, with $\delta$ chosen so the reference never overlaps the tail;
    \item the two \textit{halves} of the tail, used for an internal consistency check.
\end{itemize}
We use $f_\text{tail}=0.10$, $f_\text{ref}=0.05$, and $\delta = 0.15$.

Within each segment, the values from every realization are binned into windows of width $\Delta t = N$, yielding an ensemble of smoothed trajectories. Binning reduces temporal autocorrelations so that the resulting variance reflects genuine stochastic uncertainty across realizations.
We then compute two Geweke $Z$-scores:
\begin{align}
    Z_\text{cross} &= \frac{\mu_\text{ref} - \mu_\text{tail}}
        {\sqrt{\widehat{\sigma}^2_\text{ref}/M +
               \widehat{\sigma}^2_\text{tail}/M}}, \\[4pt]
    Z_\text{internal} &= \frac{\mu_\text{tail,lo} - \mu_\text{tail,hi}}
        {\sqrt{\widehat{\sigma}^2_\text{lo}/M +
               \widehat{\sigma}^2_\text{hi}/M}},
\end{align}
where $\mu$ denotes the mean and $\widehat{\sigma}^2$ the variance computed on the smoothed values, and ``hi" (``lo") refers to the second (first) half of the tail interval.
$Z_\mathrm{cross}$ tests whether the tail is statistically consistent with the earlier reference window, while $Z_\mathrm{internal}$ checks for residual drift within the tail itself.

Stationarity is declared if both $|Z_\text{cross}| < Z^*$ and $|Z_\text{internal}| < Z^*$, with threshold $Z^* = 2.0$.
If either condition is violated, the simulation is rerun with a larger number of realizations or a longer total time until both are satisfied.
The steady-state averages $\langle S_{\mathrm{tot}}\rangle$ are computed by averaging over the tail window and over all $M$ realizations.

\subsection{Transient times to steady state}
\label{app:transient_times}

\begin{figure}[t!]
    \includegraphics[width=\linewidth]{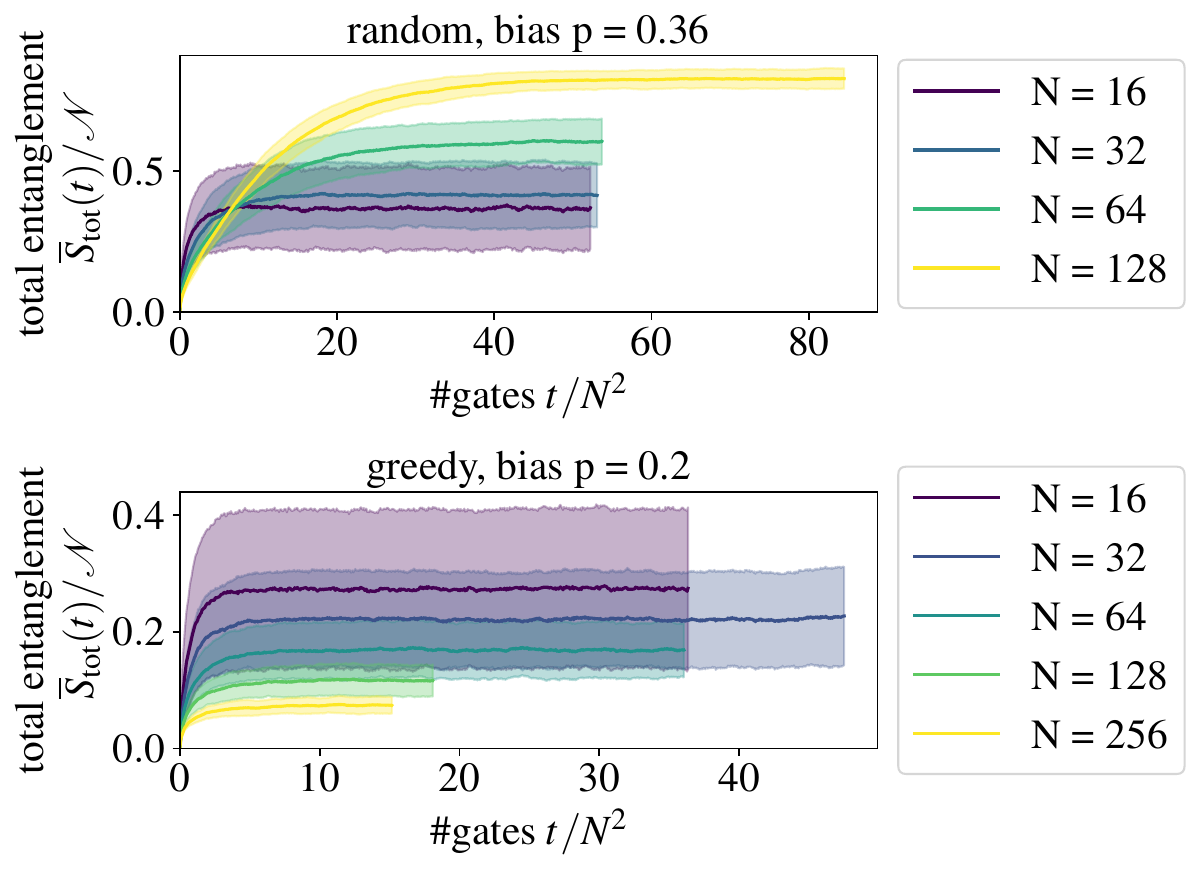}
    \caption{
    \textbf{Example smoothed trajectories $\overline{S}_\mathrm{tot}(t)$.}
    The ensemble mean (solid line) and one standard deviation (shaded band) 
    are plotted against the rescaled time $t/N^2$.
    Top panel: random strategy at $p = 0.36$.
    Bottom panel: greedy strategy at $p = 0.2$.}
    \label{fig:example_trajectories}
\end{figure}

\begin{figure}[h]
    \includegraphics[width=\linewidth]{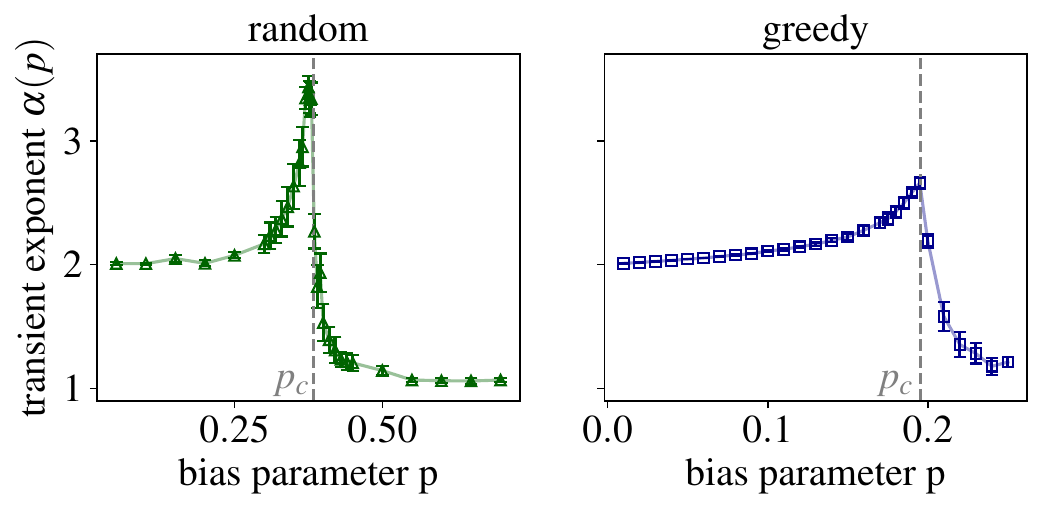}
    \caption{
    \textbf{Extracted scaling exponent $\alpha(p)$ as a function of bias parameter $p$.}
    The exponent characterize the system-size dependece of the transient time $\tau(N,p) \propto N^{\alpha(p)}$, following the fit with a single-exponential relaxation ansatz $\overline{S}_\mathrm{tot}(t) = S_\infty - A e^{-t/\tau}$.
    $p_c$ is shown as a dashed vertical line.
    In the volume-law phase ($p<p_c$), the exponent takes a value of $\alpha \approx 2$, while in the area-law phase ($p>p_c$) it falls to $\alpha \approx 1$, with a pronounced peak near the phase transition.
    Results are shown for both the random and greedy strategies.
    }
    \label{fig:transient_time_alpha}
\end{figure}

From the simulated trajectories, we estimate the time — measured in number of gates — required
for the circuit to reach the steady state under the random and greedy strategies.

Raw trajectories are smoothed by partitioning the time axis into non-overlapping bins of width $\Delta t = N$ and computing, for each bin $k$, the ensemble mean and standard deviation across all (realization, time-step) pairs falling within that bin:
\begin{equation}
    \bar{v}_k = \frac{1}{|\mathcal{B}_k|}\sum_{i \in \mathcal{B}_k} v_i,
    \qquad
    \sigma_k  = \sqrt{\frac{1}{|\mathcal{B}_k|}\sum_{i \in \mathcal{B}_k} v_i^2 - \bar{v}_k^2},
\end{equation}
where $\mathcal{B}_k$ collects all (realization, time-step) pairs whose time falls in bin $k$.
This yields a single smoothed trajectory $\{(t_k,\,\bar{v}_k)\} =: \{(t,\,\overline{S}_\mathrm{tot}(t))\}$ on a uniform grid of bin centres $t_k$. Representative examples are shown in Fig.~\ref{fig:example_trajectories}.

To extract the transient timescale $\tau(N, p)$, we fit each smoothed trajectory $\overline{S}_\mathrm{tot}(t)$ to the single-exponential relaxation ansatz
    \begin{equation}
    \overline{S}_\mathrm{tot}(t) = S_\infty - A\,e^{-t/\tau},
\end{equation}
with free parameters $S_\infty$, $A$, and $\tau > 0$.
The fit is performed via nonlinear least squares.

\begin{figure*}[!t]
    \includegraphics[width=\textwidth]{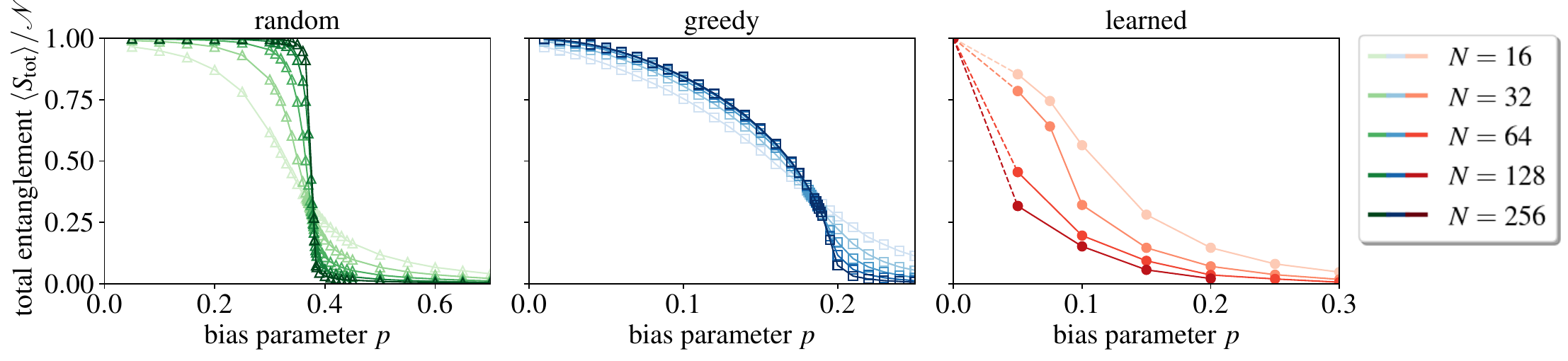}
    \caption{
    \textbf{Normalized ensemble-averaged steady-state entanglement $\langle \Stot \rangle$ versus bias parameter $p$.}
    The data from Fig.~\ref{fig:combined} are shown here, with results for $N = 256$ added for the random and greedy strategies, separated into three panels for improved readability, allowing clearer visualization of the distinct scaling behaviors across different system sizes.
    }
    \label{fig:extended_results}
\end{figure*}

To characterize how the transient time grows with system size at each measurement rate $p$, we assume a power-law scaling
\begin{equation}
    \tau(N, p) \propto N^{\alpha(p)}.
\end{equation}
The exponent $\alpha(p)$ is extracted by performing a linear regression of $\ln\tau$ against $\ln N$ across all available system sizes; the slope gives $\alpha(p)$, and its standard error provides the uncertainty estimate.
The resulting exponent $\alpha(p)$ as a function of bias parameter $p$ is shown in Fig.~\ref{fig:transient_time_alpha}.

\section{Supplementary results}
\label{app:supp_results}
In this section, we list some supplementary results that may provide useful insights for interested readers, though they do not directly impact the primary findings discussed in the main text.

\subsection{Enhanced data visualization}
\label{app:enhanced_data}
Figure~\ref{fig:extended_results} reproduces the data shown in Fig.~\ref{fig:combined} of the main text, with each strategy displayed in a separate panel. 
This layout avoids overlapping curves and makes it easier to compare scaling behaviors for each strategy individually.
In addition, the figure includes results for $N = 256$ for both the greedy and random strategies not shown in the main text.

\subsection{Finite-size scaling analysis}
\label{app:fss}
\begin{figure}[h]
    \includegraphics[width=\linewidth]{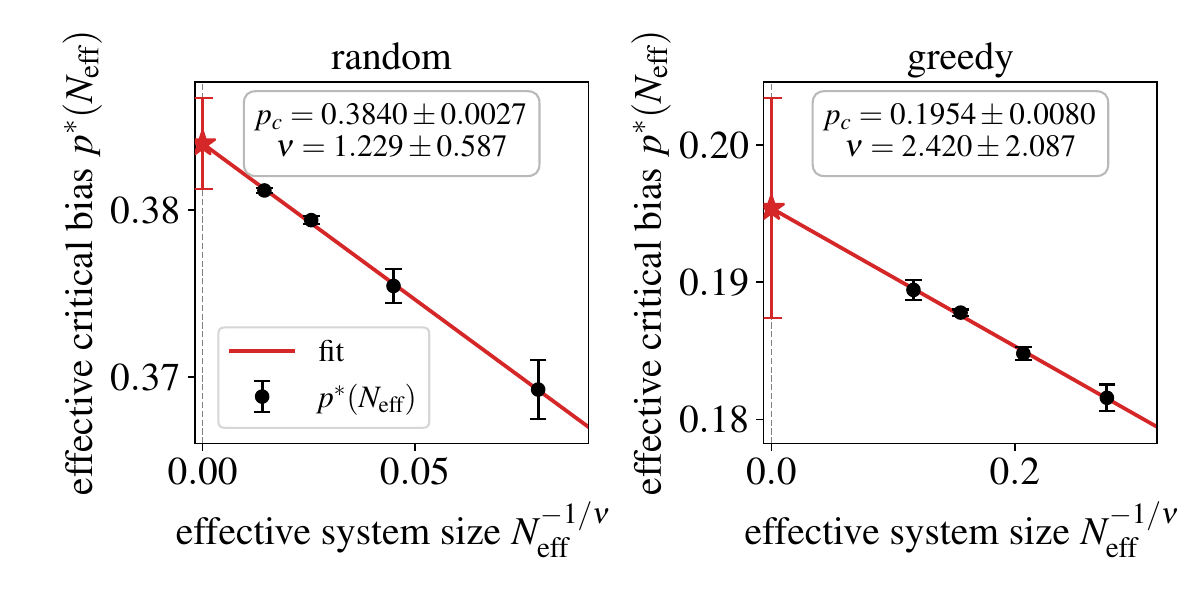}
    \caption{
    \textbf{Finite-size scaling analysis for the random and greedy strategies.}
    Effective crossing points $p^*(N_\mathrm{eff})$ plotted against $N_\mathrm{eff}^{-1/\nu}$ for the random (left) and greedy (right) strategies.
    Black points with error bars show the crossing-point estimates with $\chi^2/\mathrm{dof}$-rescaled uncertainties; the red line is the linear fit of Eq.~\eqref{eq:fss_ansatz}, and the red star marks the extrapolated critical point $p_c$ as $N_\mathrm{eff} \to \infty$.
    }
    \label{fig:fss}
\end{figure}

To extract the critical bias $p_c$ and correlation-length exponent $\nu$, we perform a finite-size scaling (FSS) analysis based on the drift of pairwise crossing points.
For each consecutive pair of system sizes $(N_1, N_2)$, we construct the difference curve
\begin{equation*}
    \Delta S(p) = \frac{\langle S_\mathrm{tot}(p,N_1)\rangle}{\mathcal{N}}
                - \frac{\langle S_\mathrm{tot}(p,N_2)\rangle}{\mathcal{N}},
\end{equation*}
evaluated on the discrete set of sampled $p$-values within a fixed fitting window, with adjacent points connected by linear segments.
The effective critical bias $p^*(N_\mathrm{eff})$ is defined as the zero of this piecewise-linear interpolant; when multiple sign changes are present, the one whose left bracket is nearest to the window centre is selected.
Under standard FSS theory, the effective critical points drift toward the thermodynamic critical point according to
\begin{equation}
    p^*(N_\mathrm{eff}) = p_c + A \, N_\mathrm{eff}^{-1/\nu},
    \label{eq:fss_ansatz}
\end{equation}
where $N_\mathrm{eff} = \sqrt{N_1 N_2}$ is the geometric mean of the two system sizes.
The three free parameters $(p_c, A, \nu)$ are determined by minimising the sum of squared residuals using differential evolution (DE).

Uncertainties on $p_c$ and $\nu$ are obtained by propagating the sampling noise through the entire pipeline: each mean $\langle S_\mathrm{tot}(p, N)\rangle/\mathcal{N}$ is independently perturbed by a zero-mean Gaussian with standard deviation equal to the sample standard deviation at that point, the crossing points are re-extracted by linear interpolation, and the model is re-fitted.
Repeating this procedure yields raw uncertainties $\sigma^\mathrm{raw}_{p_c}$ and $\sigma^\mathrm{raw}_\nu$.
In practice, we find raw uncertainties tend to underestimate the true uncertainty in $p^*$ owing to residual autocorrelations in the sampled data and the coarse spacing of the $p$-grid. 
When the fit residuals exceed the bootstrap error bars, we apply a $\chi^2/\mathrm{dof}$ rescaling.
Specifically, we compute
\begin{equation}
    \frac{\chi^2}{\mathrm{dof}}
    = \frac{1}{N_\mathrm{pairs} - 3}
      \sum_i \left(\frac{p^*_i - \hat{p}^*_i}
                        {\sigma^{\mathrm{raw}}_{p^*_i}}\right)^2,
\end{equation}
where $\hat{p}^*_i = p_c + A\, N_{\mathrm{eff}, i}^{-1/\nu}$ are the model predictions and $\sigma^{\mathrm{raw}}_{p^*_i}$ are the raw bootstrap standard deviations of the individual crossing points.
All reported uncertainties are then rescaled by a factor $s = \sqrt{\max(1,\,\chi^2/\mathrm{dof})}$, giving $\sigma_{p_c} = s\,\sigma^\mathrm{raw}_{p_c}$, and likewise for $\nu$.

The results of the FSS analysis are shown in Fig.~\ref{fig:fss} for random and greedy strategies.
The extracted critical points and exponents are: random strategy
$p_c = 0.384 \pm 0.003$, $\nu = 1.2 \pm 0.6$; greedy strategy
$p_c = 0.195 \pm 0.008$, $\nu = 2.4 \pm 2.1$.

\subsection{Beyond the greedy strategy: a greedier variant}
\label{app:greedier}

\begin{figure}[t]
    \includegraphics[width=\linewidth]{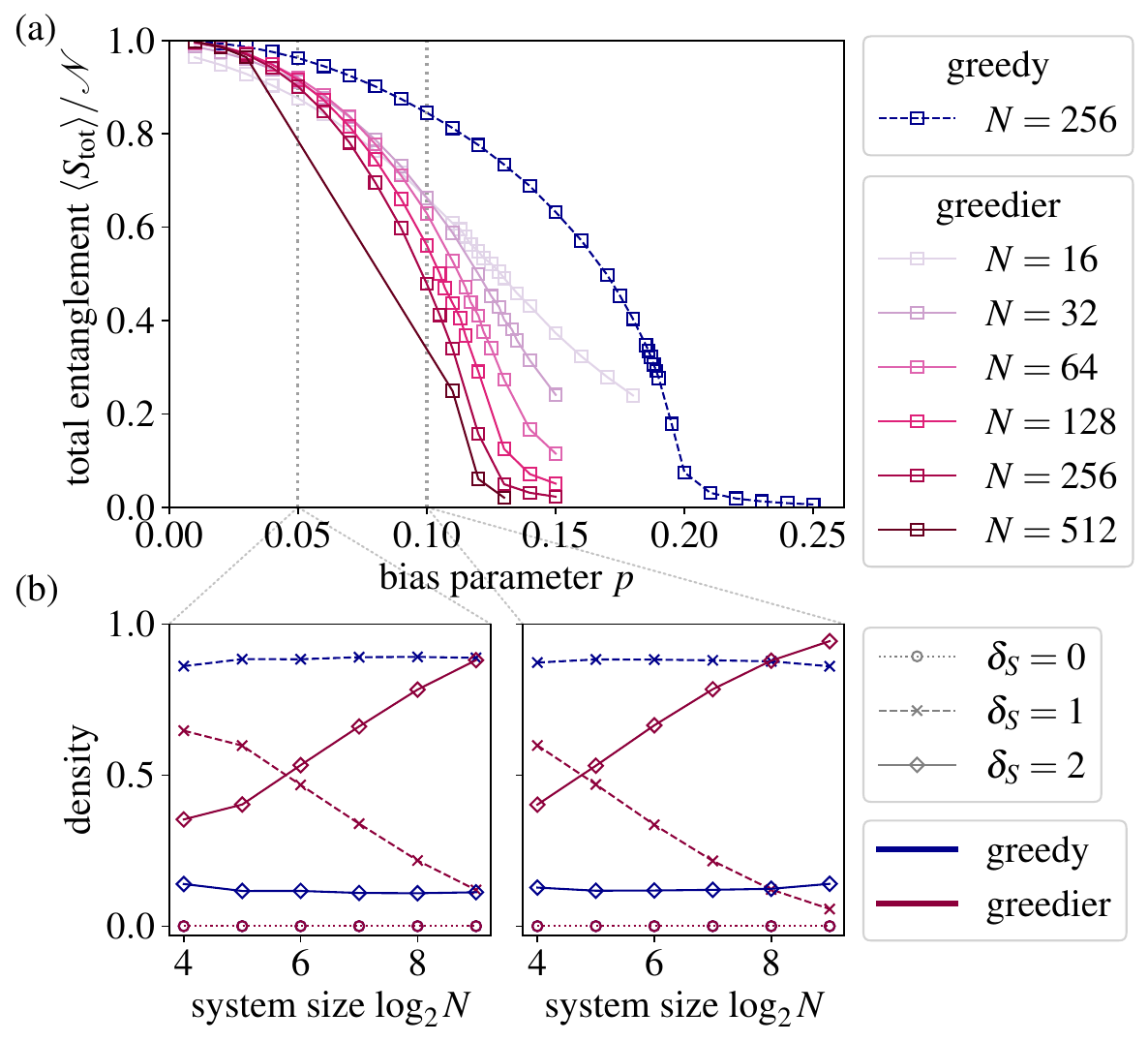}
    \caption{
    \textbf{Greedy vs.\ greedier strategy comparison.}
    (a) Normalized average steady-state entanglement $\langle S_\mathrm{tot} \rangle$ for the greedier strategy at different system sizes (colored squares and solid lines). The blue dashed line with squares shows the same quantity for the greedy strategy at $N=256$.
    (b) Density of disentangling gates removing $\delta_S = 0, 1, 2$ units of entanglement in the steady state, shown for different system sizes and two values of the bias parameter: $p=0.05$ (left) and $p=0.1$ (right). 
    The greedy strategy shows similar densities across different system sizes, whereas the changes for the greedier strategy highlight the presence of significant finite-size effects.
    }
    \label{fig:greedier}
\end{figure}

The greedy strategy evaluates all possible gate locations and selects uniformly at random among those locations $a$ for which $\Delta S(a) > 0$, where $\Delta S(a) := S_t(a) - S_{t+1}(a)$ denotes the reduction in entanglement produced by a disentangling gate at location $a$.

A natural refinement is to also maximize over $\Delta S(a)$. For Clifford circuits, the possible values are $\Delta S(a)/\ln 2 \in \{0, 1, 2\}$, so a \textit{greedier} strategy selects among those locations achieving the maximum reduction $\Delta S_\mathrm{max} := \max_a \Delta S(a)$, breaking ties uniformly at random:
\begin{equation}
    \pi_{\mathrm{greedier}}(a \mid s) = 
    \begin{cases}
        \dfrac{1}{Z}, & \text{if } \Delta S(a) = \Delta S_\mathrm{max}, \\[6pt]
        0, & \text{otherwise,}
    \end{cases}
    \label{eq:greedier_policy}
\end{equation}
where $Z = |\{a : \Delta S(a) = \Delta S_\mathrm{max}\}|$ is the number of maximally disentangling locations; the dependence of $\Delta S(a)$ on the state $s$ is left implicit.

We measured the steady-state value of $\langle S_\mathrm{tot} \rangle$ for this greedier strategy following the same procedure used for the random and greedy strategies (see App.~\ref{app:sampling_steady_state}). The results are shown in Fig.~\ref{fig:greedier} (a), alongside the greedy strategy at $N=256$ for reference. The greedier strategy is clearly more efficient: by placing gates that remove $\Delta S = 2$ units of entanglement more frequently, it stabilizes an area-law phase in a regime where the greedy strategy exhibits volume-law entanglement.

From the top panel's data, the greedier strategy does not appear to have a well-defined critical point $p_c$, suggesting the absence of a volume-law phase in the thermodynamic limit. However, a more careful analysis reveals that this is an artifact of finite-size effects, as we now demonstrate.

In the steady state, we compute the relative densities of gates removing $\delta_S = 0, 1, 2$ units of entanglement, and plot them as a function of system size for $p = 0.05$ and $p = 0.1$, Fig.~\ref{fig:greedier} (b). 
First, the $\delta_S = 0$ density is zero for both strategies: they always place disentangling gates. 
Second, while the greedy strategy shows a stable distribution across system sizes, the greedier strategy exhibits a strong size dependence up to $N = 512$. This occurs because locations with $\delta_S = 2$ are rarer than those with $\delta_S = 1$, so the greedier strategy can exploit its advantage only at large system sizes ($N \gtrsim 1024$), where such locations become sufficiently abundant. 
A rigorous characterization of the thermodynamic behavior would therefore require numerically prohibitive system sizes, particularly given the slow thermalization times discussed in Sec .~\ref{app:transient_times}.

Finally, we note a simple consistency check: since the greedier strategy removes roughly twice as much entanglement per step as the greedy strategy, one expects $p_c^\mathrm{greedy} \simeq 2\, p_c^\mathrm{greedier}$. The top panel of Fig.~\ref{fig:greedier} suggests $p_c^\mathrm{greedier} \simeq 0.12$ for $N = 512$, which is consistent with $p_c^\mathrm{greedy} = 0.195 \pm 0.008$.

\subsection{Stability of human-designed strategies}
\label{app:human_designed_strategies}
\begin{figure}[t]
    \includegraphics[width=\linewidth]{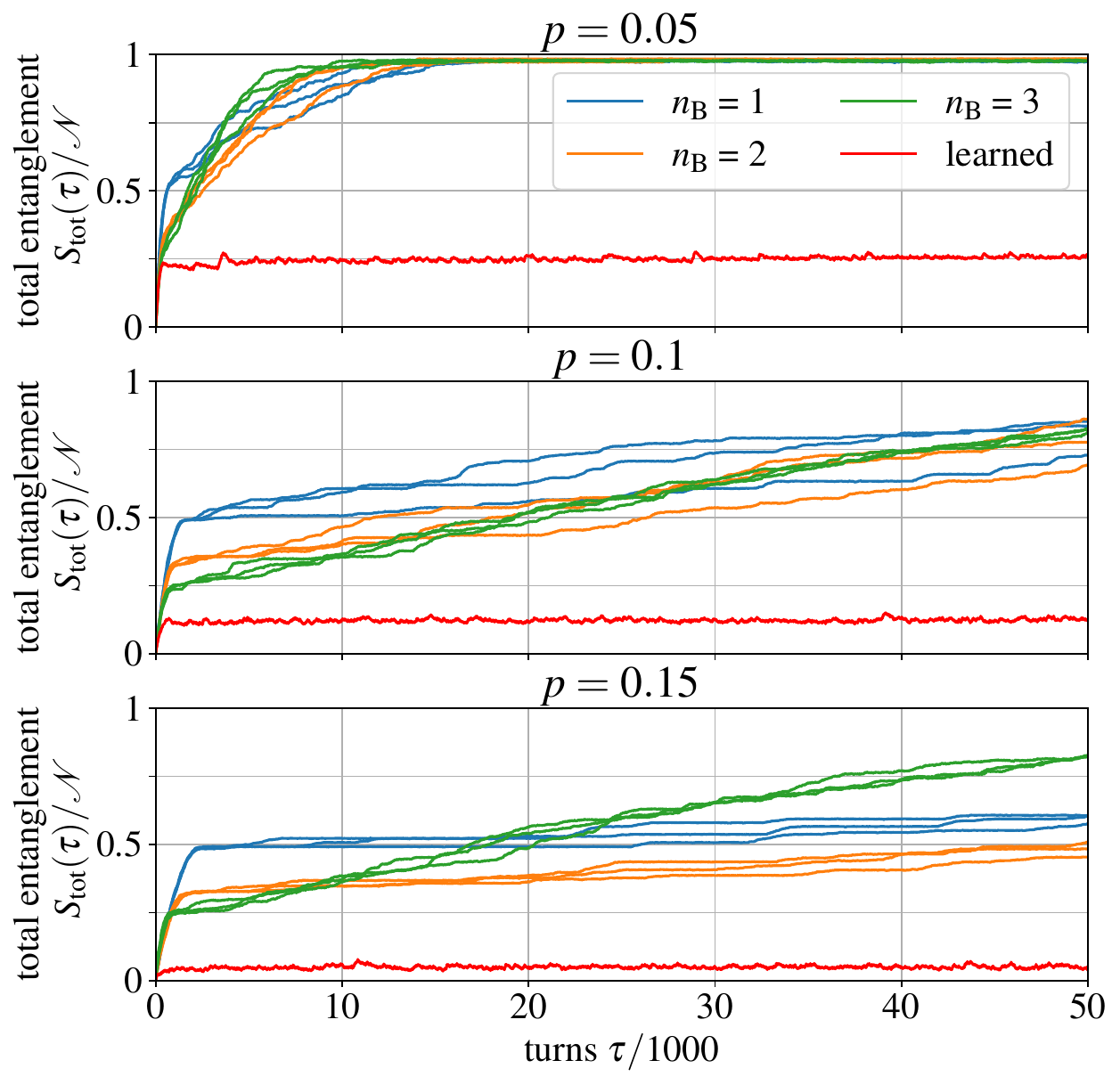}
    \caption{
    \textbf{Total entanglement $\Stot (\tau)$ dynamics under learned and pyramid human-designed strategy.} 
    The three plots display the entanglement dynamics $\Stot (\tau)$ starting from the product state $\ps$ for a system size $N=128$ and $p=0.05,0.1,0.15$.
    Red lines represent a single realization of the learned strategy.
    Color-coded lines represent realizations of the pyramid human-designed strategy with different values of the number of bottlenecks $n_\mathrm{B}$ (three realizations per $n_\mathrm{B}$ value); see text for details.
    The data show that the learned strategy reaches a steady state with finite average $\Stot$, whereas under the pyramid strategy $\Stot$ keeps growing well beyond the value of the learned strategy, saturating close to its maximal value for $p=0.05$ within the simulated time window.
    The growth is slower for larger $p$.
    This demonstrates that the pyramid strategy is not successful in sustaining entanglement bottlenecks, i.e., pyramidal structures in the entanglement profile.
    }
    \label{fig:pyramids_human}
\end{figure}
In Section~\ref{sec:informed_strategies}, we noted that learned strategies, while appearing simple, are difficult to replicate through a simple set of rules.
To showcase this, we consider two strategies -- a simple and a more sophisticated one -- and study their steady state.

We start with what we call ``pyramid strategy''.
It implements the intuition that it suffices to act on the bottlenecks and their neighborhood.
For a fixed number $n_\mathrm{B}$ of bottlenecks, the strategy targets the equidistant bonds $b_i = i \lfloor (N-1)/(n_\mathrm{B}+1) \rfloor$, $i=1,\dots,n_\mathrm{B}$, together with their neighboring bonds $b_i \pm 1$ (three bonds per bottleneck).
At each turn, it acts on a bottleneck bond $b_i$ if any of them admits a disentangling gate, otherwise on a left neighbor $b_i - 1$, and otherwise on a right neighbor $b_i + 1$; within each group, it selects the bond with the largest entanglement reduction $\Delta S$ (the leftmost one in case of ties).
If none of the $3 n_\mathrm{B}$ target bonds admits a disentangling gate, the strategy acts on $b_1 + 1$, which leaves the entanglement profile unchanged, i.e., it effectively passes its turn.
We analyze the dynamics for system size $N=128$ -- the largest size for which a learned strategy is available, where finite-size effects are weakest, and where pyramids emerge most clearly -- for $p=0.05,0.1,0.15$. 
We focus on single realizations of the dynamics, sufficient to identify the strategy's failure.
Fig.~\ref{fig:pyramids_human} presents numerical results with a single realization of the learned strategy (red lines) and three different realizations per parameter $n_\mathrm{B}=1,2,3$ (blue, orange, green lines).
The pyramid strategy features an initial transient phase with rapid entanglement growth -- especially at low bias ($p = 0.05, 0.1$) and with many bottlenecks -- stabilizing at higher $\Stot$ values than learned strategies.
This configuration is not stable.
The stochastic evolution acting at and next to a bottleneck can bring all three target bonds into configurations where none of them admits a disentangling gate -- for instance, when they lie on a segment of maximal slope of the entanglement profile, or at a local minimum.
The bottlenecks are lost and the steady state saturates at the maximum value of entanglement. \\

\begin{figure}[t]
    \includegraphics[width=\linewidth]{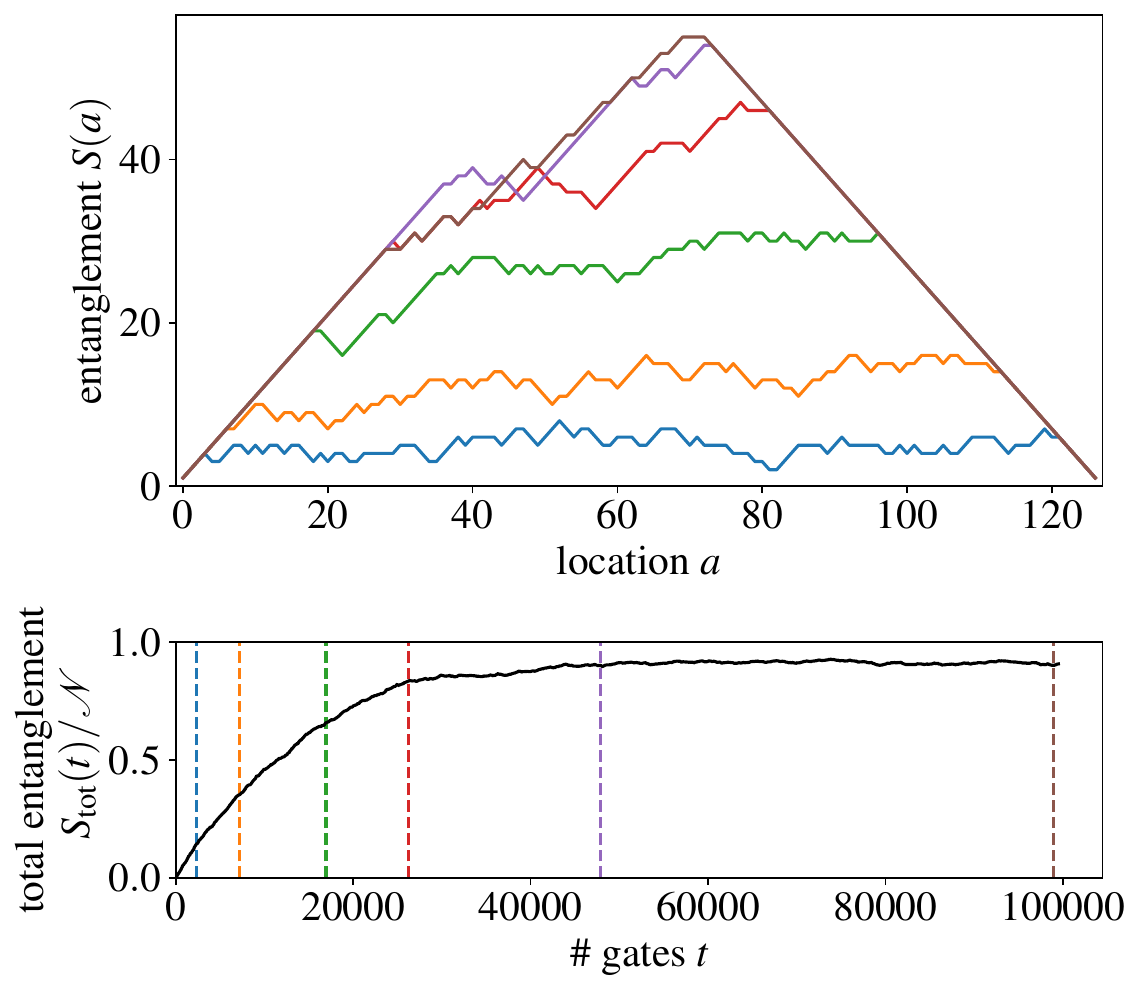}
    \caption{
    \textbf{Entanglement profile and $\Stot (\tau)$ dynamics under smart human-designed strategy.} 
    Dynamics of a single circuit realization starting from a product state $\ket{0}^{\otimes N}$ for a system size $N=128$ and $p=0.1$.
    Top: entanglement profiles at selected times, color-coded according to time.
    Bottom: dynamics of $S_{\rm tot}(t)$, with time $t$ measured in the number of applied gates. Vertical dashed lines indicate the times at which the corresponding entanglement profiles are shown in the top panel.
    The smart strategy fails to sustain the pyramidal structure in the entanglement profile.
    }
    \label{fig:smart-strategy}
\end{figure}
One may argue that the pyramid strategy is too simple. 
We therefore present here a more sophisticated strategy, suggested by one of the referees during the review process.

This ``smart strategy'' builds on the following observation: what matters is not the bottlenecks per se, but that they induce extended inactive regions of maximal steepness (in the entanglement profile) where neither the entangler nor the disentangler can change the entanglement, so that the effective rate of consequential entangling moves is reduced.
The strategy takes the entanglement profile as input.
We identify three types of bonds:
\begin{itemize}
    \item e-active bond: where the entanglement can be increased.
    \item bottlenecks: bonds with zero entanglement.
    \item sinks: bonds with non-zero entanglement and neighbouring bonds (before and after) with higher entanglement -- a local minimum.
\end{itemize}
The goal of the strategy is to reduce the number of e-active bonds. 
The strategy is:
\begin{enumerate}
    \item If there exists a bond such that disentangling yields $\Delta S = 2$ and reduces the number of e-active bonds, take it. In case of degeneracy, start with the leftmost one.
    \item Else, act on the bond that is closest to a sink that does not have a closer bottleneck. 
    The idea is that a bond of this type will have an e-active bond next to it, and placing a disentangling gate here will move that e-active bond.
    Moving the e-active bond towards a sink will create a region where step 1 is feasible; instead, if we move the e-active bond towards a bottleneck, we would not be able to remove it. 
    If there are multiple options, take the right-most\footnote{Left-most would also work, but choosing right-most appears to empirically give more uniform steady states.}.
    \item Else, the best disentangling move increases the number of e-active bonds (it is a peak in the entanglement profile); then pick the bond closest to an existing sink. 
    \item Otherwise, choose the left-most bond.
\end{enumerate}

We evaluated this strategy at $N=128$, $p=0.1$.
In Fig.~\ref{fig:smart-strategy}, we report the entanglement profile at selected times (top) and the dynamics of $S_{\rm tot}(t)$ (bottom).
The data show that this strategy cannot sustain the pyramidal structure in the entanglement profile either: the total entanglement saturates at a near-maximal value, $\Stot/\mathcal{N} \approx 0.9$.
Once more, the simple rules we tested do not reproduce the learned strategies.

We would like to note that the entanglement profile is only an \emph{effective} description of the Clifford state: many distinct stabilizer states share the same profile, and moves that are legal in the classical height model need not correspond to admissible operations on the underlying state. 
As a simple illustration, consider a four-qubit state with entanglement $S_i = \ln 2$ across all three cuts $i=0,1,2$. 
Several inequivalent stabilizer states are compatible with this profile, and they require different orderings of the disentangling gates to be brought to a product state. 
A rule formulated purely at the level of the heights therefore cannot, in general, distinguish situations that demand different actions.

\subsection{Learned strategies with vanishing information}
\label{app:vanishing_info}
\begin{figure}[t]
    \includegraphics[width=\linewidth]{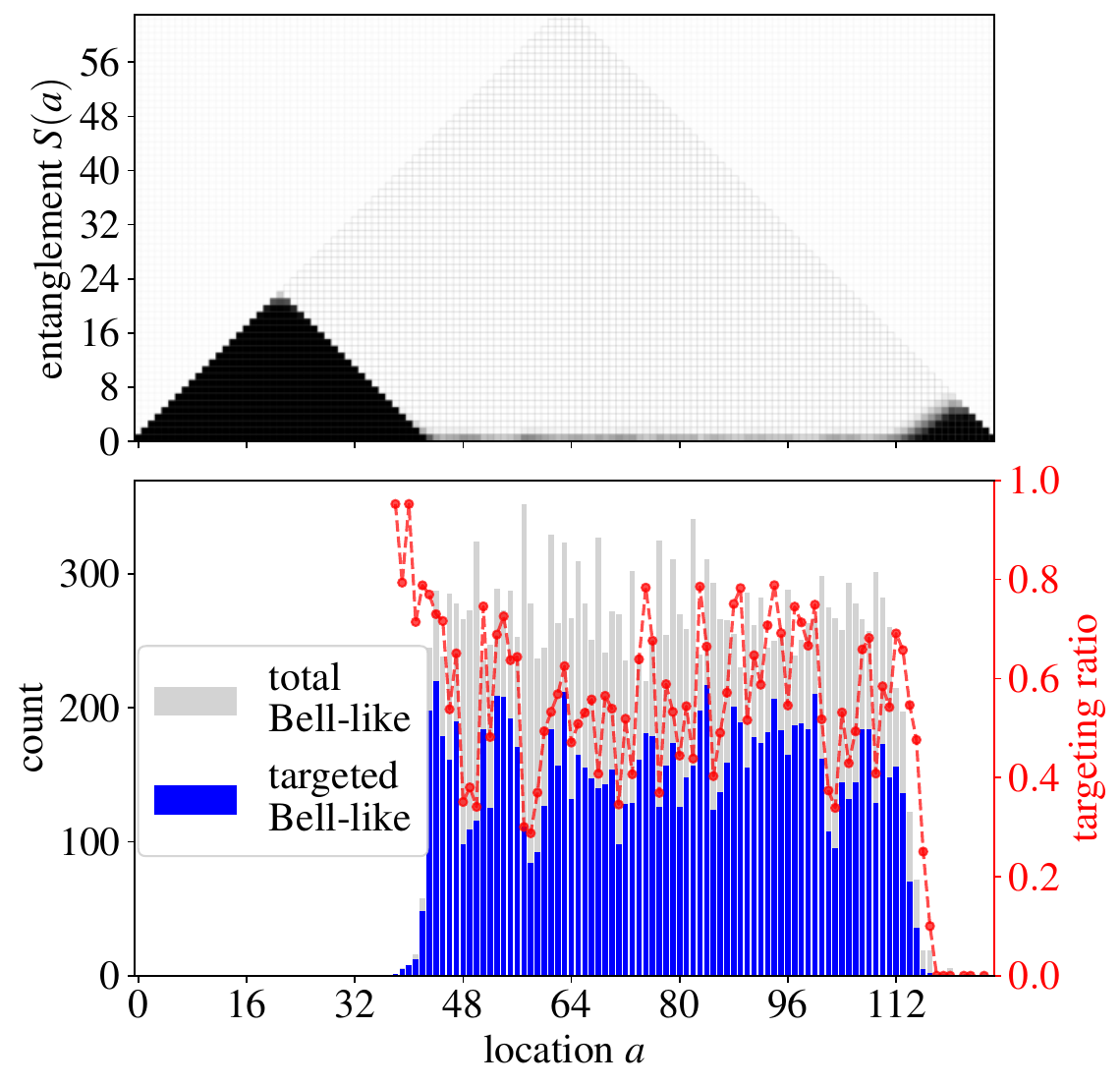}
    \caption{
    \textbf{Bell pair targeting for learned model.}
    The top panel shows the entanglement spectrum averaged over 30,000 turns $\tau$. 
    Each square represents the density of entanglement averaged over time, encoded in grayscale.
    The plot shows that the steady state is stable over time, with only minor fluctuations visible near the center of the chain.
    The bottom panel illustrates how the learned model targets Bell-like pairs. 
    The gray histogram indicates the total number of Bell-like pairs in the observations during the dynamics, while the blue histogram shows how often the model actively targeted them. 
    The red line represents the (rescaled) targeting ratio, defined as the number of targeted Bell-like pairs divided by the total number observed. 
    The data are obtained from the learned model at $N=128,p=0.2,q=0.1$.
    }
    \label{fig:vanishing_info}
\end{figure}
In Section~\ref{sec:partial_info}, we conjecture that for $p > \pc^{\mathrm{greedy}}$, area-law scaling persists as $q \to 0$.
Here, we present numerical evidence supporting this conjecture.

The underlying intuition is that, as the system size grows, the gate density per bond decreases, and entanglement arises mainly from nearest-neighbor Bell-like pairs -- since only nearest-neighbor gates are applied.  
In the vanishing-information limit ($q \to 0$), each observation consists of a single stabilizer providing localized information about one Bell-like pair. 
This information can always be exploited to eliminate the pair before it spreads across larger regions of the system. In this way, the learned strategy effectively operates in a greedy fashion, targeting Bell-like pairs.

To support our conjecture, we investigate how learned strategies behave in the regime $p > \pc^\mathrm{greedy}$; in particular, we focus on the case $N=128$, $p=0.2$, and $q=0.1$. 
We are interested in characterizing how the strategy targets Bell-like pairs in the steady state. The data are presented in Fig.~\ref{fig:vanishing_info}.

The upper panel shows the entanglement spectrum averaged over 30{,}000 turns $\tau$, where grayscale encodes frequency. 
The profile is stable in time, with low entanglement maintained in the central region between the two pyramids. 
These pyramid-like structures are likely due to finite-size effects and are expected to vanish in the thermodynamic limit.

The lower panel provides statistics on Bell-like pairs observed and targeted during the 30{,}000 turns in the steady state. 
For each turn $\tau$, we extract from the observation the number and location of Bell-like pairs. 
The gray histogram shows the counting of Bell-like pairs across the trajectory, while the blue histogram shows how often the strategy actively targets them. 
The red curve represents the rescaled targeting ratio, defined as the number of targeted Bell-like pairs divided by the total number observed.
The data highlights that the strategy frequently targets Bell-like pairs in the low-entanglement region, consistent with the conjectured behavior.
Note that the blue histogram does not fully overlap the gray one. Since $q=0.1$ implies an average of $\sim13$ stabilizers per observation, multiple Bell-like pairs may be present in a single turn. However, the model can place at most one disentangling gate per turn, limiting the targeting capacity.
However, the conditional targeting probability is $76\%$.%

The targeting ratio (red line) shows a bias toward the left region of the system. 
The emergence of pyramid structures close to the boundaries is attributed to finite-size effects. 
The strategy focuses on this region to prevent the growth of the pyramid, which would otherwise lead to a maximally entangled state.

These results suggest that the learned strategies are capable of maintaining area-law entanglement scaling by persistently acting on Bell-like structures whenever present in the observations. 

\section{Reinforcement learning methodology}
\label{app:rl}
This section outlines the technical aspects of the RL pipeline. 
The entire implementation is written in Python using the JAX library~\cite{jax2018github}, which enables end-to-end GPU acceleration for high-performance computation. 
The code, trained models, and the hyperparameters are available in the Zenodo repository~\cite{qtetrix_repo}.

\subsection{Environment setup and dynamics}
\label{app:rl_environment}
To learn and investigate entanglement-reducing strategies in Clifford circuits (Sec.~\ref{sec:model}), we implement a custom RL environment. 
The RL pipeline, outlined in Sec.~\ref{sec:rl_circuit_control} and Fig.~\ref{fig:rl_scheme}, trains an agent to learn a policy $\pi(a|s)$ that selects positions $a$ along a one-dimensional chain where disentangling two-qubit gates are applied. 
At each step, the environment first places a disentangling gate at the selected position $a$, then updates the tableau description by simulating stochastic circuit dynamics: a random number $n_e$ of two-qubit Clifford gates is applied, with $n_e$ sampled from the geometric distribution $(1-p)^{n_e}$ and each gate drawn uniformly from $\mathcal{C}_2$.

The RL environment, following the standard RL framework, has the following components:
\begin{itemize}
    \item \textbf{State}: The internal state is represented by a (unphased) stabilizer tableau $T$, an $N \times 2N$ binary $(0,1)$ matrix where $N$ is the system size. Each row encodes a stabilizer generator in the $X_i, Z_i$ basis without phase information (see App.~\ref{app:stabilizer_circuits}).
    \item \textbf{Observation}: The observation fed back to the agent is the tableau encoded as $(1,-1)$. 
    Partial observability consists of setting each stabilizer (tableau's row) to zero with probability $1-q$
    This encoding choice serves two purposes: neural networks perform better with normalized, zero-centered data, and it allows distinguishing between trivial stabilizer actions ($-1$) and missing information ($0$) when $q < 1$.
    \item \textbf{Action}: The action space consists of integers $a \in [0, N-2]$ specifying the location along the chain where a disentangling gate is applied.
    \item \textbf{Reward}: The reward function measures the total entanglement reduction across all bonds, normalized to $[0,1]$: 
    $$r = -\frac{\Stot}{\mathcal{N}} = -\frac{1}{\mathcal{N}} \sum_i S(i),$$
    where $S(i)$ is the entanglement entropy across the bipartition $\{1,\dots,i\} \{i+1,\dots,N\}$, and $\mathcal{N}$ is the normalization constant defined in \eqref{eq:normalization} (see also below).
    \item \textbf{Termination}: Episodes terminate when the entanglement entropy reaches zero across all bonds, indicating complete disentanglement of the system, or after reaching the maximum number of steps \texttt{episode\_length}.
\end{itemize}
The normalization factor $\mathcal{N}$ corresponds to the maximum possible total entanglement $\Stot$ in the system. 
It is obtained by summing the maximal entanglement entropy $S_{\mathrm{max}}(i) = \min\{i+1, N-1-i\}$ across all bipartitions, which yields a symmetric, pyramidal profile over the $N-1$ bonds of the chain (see Fig.~\ref{fig:rl_fullinfo}(a)). 
For a general base of length $x$, the total area of this profile is given by
\begin{equation}
    \mathcal{N}(x) = 
    \begin{cases}
        \left\lfloor \dfrac{x}{2} \right\rfloor \cdot \left( \left\lfloor \dfrac{x}{2} \right\rfloor + 1 \right), & \text{if } x \text{ is even} \\[8pt]
        \left( \left\lfloor \dfrac{x}{2} \right\rfloor + 1 \right)^2 , & \text{if } x \text{ is odd.}
    \end{cases}
\end{equation}
In our setting, $x = N - 1$ since there are $N - 1$ entanglement cuts (bonds) in a chain of $N$ sites.

Training an RL agent typically requires exploring many environment trajectories, i.e., simulating numerous timesteps.
Therefore, fast execution is essential.
To this end, we introduce a few design choices that significantly improve runtime performance:
\begin{itemize}
    \item We precompute all disentangling gates and store them in a dictionary, each associated with a unique identifier.
    This identifier is computed from the relevant stabilizers that describe the two qubits to be disentangled.
    As a result, the disentangling step reduces to a dictionary lookup, avoiding the need to solve a minimization problem (see App.~\ref{app:disentangling_cliffords}).
    \item Similarly, we pre-generate the elements of the two-qubit Clifford group $\mathcal{C}_2$ and store them in an array.  
    Randomly sampling from this array is more efficient than generating a new random gate at every timestep.
    \item The initial states are precomputed and stored in an array during initialization.  
    These can be either the product state $\ps$ or randomly generated states, obtained by applying random circuits of fixed depth \texttt{n\_init\_gates}.  
    A total of \texttt{n\_init\_random\_states} such initial conditions are cached for fast reuse.
\end{itemize}

Following the standard RL environment design, the two core methods are \texttt{reset} and \texttt{step}. 
The \texttt{reset} method initializes the tableau either to the identity (corresponding to the product state $\ps$) or to a randomly sampled configuration from the cached random initial conditions. 
The \texttt{step} method applies the disentangling gate at location $a$ specified by the agent, followed by the stochastic dynamics. 
Pseudocode for this method is provided in Algorithm~\ref{alg:step}. 
To aid understanding, we clarify the main components:
\begin{itemize}
    \item The array \texttt{disentangling\_gates} stores all available disentangling gates, and \texttt{delta\_entanglement} contains their associated entanglement changes.
    \item \texttt{compute\_id} takes the current tableau and action as input, returning a unique identifier \texttt{id}.
    \item \texttt{apply\_gate} applies the selected two-qubit gate to the tableau via matrix multiplication on the relevant subblock.
    \item \texttt{clip\_tableau} applies the clipping algorithm (see App.~\ref{app:clipping_algorithm}), which fixes the gauge and computes the entanglement across all bipartitions.
    \item \texttt{random\_choice} uniformly samples from the array at input.
\end{itemize}

\begin{algorithm}[H]
\caption{Environment step Function}
\label{alg:step}
\begin{algorithmic}
    \State \textbf{Input:} tableau, entanglement, a
    \State \textbf{Output:} updated tableau, entanglement, reward, done

    \State ID $\gets$ \texttt{compute\_id}(tableau, a)
    \State gate $\gets$ disentangling\_gates[ID]
    \State $\Delta S \gets$ delta\_entanglements[ID]

    \State tableau $\gets$ \texttt{apply\_gate}(tableau, gate, a)
    \State entanglement[a] $\gets$ entanglement[a] + $\Delta S$
    \State reward $\gets$ -\texttt{sum}(entanglement) / $\mathcal{N}$
    \State done $\gets$ (reward == 0)

    \State $r \sim U(0,1)$ \Comment{Sample from uniform distribution.}
    \While{$r > p$}
        \State gate $\gets$ \texttt{random\_choice}(random\_gates)
        \State a $\gets$ \texttt{randint}(0,N-1)
        \State tableau $\gets$ \texttt{apply\_gate}(tableau, gate, a)
        \State $r \sim U(0,1)$
    \EndWhile
    \State tableau, entanglement $\gets$ \texttt{clip\_tableau}(tableau)
\end{algorithmic}
\end{algorithm}

\subsection{Training algorithm}
\label{app:rl_ppo}
To train the RL agent, we use the Proximal Policy Optimization (PPO) algorithm~\cite{schulman2017proximal}, a state-of-the-art policy gradient method.
Policy gradient methods directly learn a parameterized policy $\pi_\theta$ by updating the parameters $\theta$ to maximize expected rewards. 
The PPO update prevents the updated policy from deviating too far from the previous one by clipping the objective function, which improves learning stability and avoids performance collapse.
This makes PPO well-suited for noisy stochastic environments such as the environment investigated in this work.
Our implementation is based on the Brax project~\cite{brax2021github} and is written entirely in JAX, enabling efficient end-to-end training on GPU.

The PPO algorithm alternates between data collection and policy updates.
During data collection, the current policy $\pi_\theta$ interacts with the environment to generate trajectories. 
\[
\mathcal{T} = \{s_t, a_t, r_t\}, \quad t = 1, \dots, \texttt{unroll\_length},
\]
where $s_t$ is the environment state, $a_t$ the action, and $r_t$ the reward.
To accelerate data collection, we evolve \texttt{num\_envs} environments in parallel, producing a batch of trajectories at each rollout step.

The collected trajectories are used to compute the objective $L(s,a,r,\theta,\theta_k)$ and update the policy parameters:
\begin{equation*}
    \theta_{k+1} = \arg\max_\theta \, \mathbb{E}_{(s,a,r) \sim \mathcal{T}} \left[ L(s,a,r,\theta,\theta_k) \right].
\end{equation*}
Policy updates are performed using the Adam optimizer with learning rate \texttt{learning\_rate}. For each batch of collected data, \texttt{num\_updates\_per\_batch} gradient updates are performed; in each update, the data is split into \texttt{num\_minibatches} of size \texttt{batch\_size}.

The overall PPO objective includes three terms:
\begin{equation*}
    L(\theta) = L^{\text{CLIP}}(\theta) - c_1 L^{\text{VF}}(\theta) + c_2 S[\pi_\theta],
\end{equation*}
where $L^{\text{CLIP}}$ is the clipped surrogate policy loss, $L^{\text{VF}}$ is the value function loss, and $S[\pi_\theta]$ is the policy entropy. The constants $c_1$ (\texttt{value\_cost}) and $c_2$ (\texttt{entropy\_cost}) control the relative weight of each term.

The value function loss $L^{\text{VF}}$ is defined as the mean squared error between the predicted state values $V_\theta(s_t)$ and the empirical returns. The value function, also known as the critic, provides a baseline estimate of expected return from a given state, which helps reduce the variance of the policy gradient estimate. Minimizing this loss improves the accuracy of the critic, leading to more stable and sample-efficient updates.

The clipped surrogate objective is defined as:
\begin{equation*}
    L^{\text{CLIP}}(\theta) = \mathbb{E}_t \left[ 
        \min\left( \rho_t(\theta) \hat{A}_t,\, \text{clip}(\rho_t(\theta), 1 - \epsilon, 1 + \epsilon) \hat{A}_t \right)
    \right],
\end{equation*}
where 
\[
\rho_t(\theta) = \frac{\pi_\theta(a_t \mid s_t)}{\pi_{\theta_k}(a_t \mid s_t)}
\]
is the probability ratio between the current and previous policy, and $\hat{A}_t$ is an estimator of the advantage function. The clipping prevents large updates by constraining $\rho_t(\theta)$ within $[1 - \epsilon, 1 + \epsilon]$ (\texttt{clipping\_epsilon}).

The advantage function is estimated via Generalized Advantage Estimation (GAE), which computes a discounted sum of temporal difference (TD) errors:
\begin{equation*}
    \hat{A}_t = \sum_{l=0}^{\infty} (\gamma \lambda)^l \delta_{t+l}, \quad \text{with} \quad
    \delta_t = r_t + \gamma V(s_{t+1}) - V(s_t).
\end{equation*}
Here, $\gamma$ is the discount factor (\texttt{discounting}) and $\lambda$ controls the bias-variance tradeoff (\texttt{gae\_lambda}). In practice, the sum doesn't run to infinity but is truncated at \texttt{unroll\_length}.
To maintain finite advantage estimates as the horizon increases, the reward is rescaled by a constant factor \texttt{reward\_scaling}.
To reduce variance during optimization, we normalize the advantages within each minibatch. 

\subsection{Policy architecture}
\label{app:rl_policy}

\begin{figure}[t!]
    \includegraphics[width=0.9\linewidth]{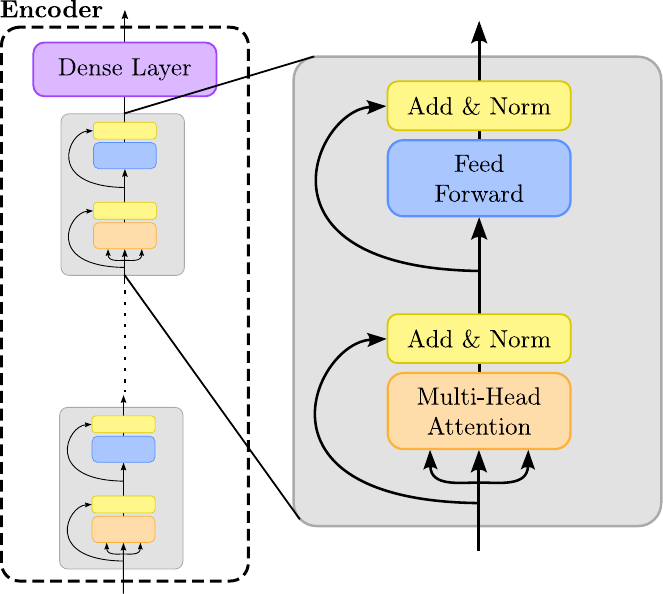}
    \caption{
    \textbf{Encoder architecture schematic.} 
    Schematic of the encoder architecture used for both actor and critic networks. 
    The encoder (left) consists of $N_\mathrm{layers}$ stacked layers, each following the structure shown on the right.
    A final dense output layer maps the representation to the desired dimensions.
    In each layer, the input is replicated to form queries, keys, and values for the multi-head attention block. 
    The output of this block is added to the input (residual connection) and passed through a normalization layer. 
    This is followed by a feedforward neural network with another residual connection and normalization. 
    }
    \label{fig:encoder}
\end{figure}
Both the actor and critic are modeled with a transformer-based architecture implemented using the Flax Python library. 
Specifically, we employ the encoder component of the transformer architecture illustrated in Fig.~\ref{fig:encoder}. 
The encoder architecture stacks $N_\mathrm{layers}$ identical layers (specified by the hyperparameters \texttt{policy\_num\_layers} and \texttt{value\_num\_layers}), where each layer follows the structure depicted on the right side of Fig.~\ref{fig:encoder}. 
After processing through all encoder layers, the data is flattened and passed through a final dense layer to produce the desired output dimension.

Each encoder block consists of four main components. First, a multi-head self-attention mechanism employs several parallel attention heads (\texttt{policy\_num\_heads} and \texttt{value\_num\_heads}), followed by a residual connection and layer normalization, where the output of the attention sublayer is added to its input and then normalized. 
Next, a feed-forward network is applied, using a GELU activation and hidden dimensions set by \texttt{policy\_feedforward\_sizes} and \texttt{value\_feedforward\_sizes}. Finally, a second residual connection and layer normalization wrap the feed-forward sublayer.

The choice of transformer architecture over simpler alternatives such as multilayer perceptrons (MLPs) or convolutional neural networks (CNNs) was motivated by both theoretical and empirical considerations. 
The tableau, input to our agent, encodes in each row a stabilizer generator. Crucially, the quantum state representation is invariant to permutations of these stabilizers -- reordering the rows of the tableau does not change the quantum state. 
Transformer architectures naturally implement this permutation invariance through their self-attention mechanism.
Furthermore, highly entangled quantum states often exhibit long-range correlations between distant sites along the chain; CNNs are inherently limited in capturing such non-local dependencies. 
In contrast, the self-attention mechanism can directly capture correlations between any pair of positions in the input sequence, making it well-suited for processing highly entangled states where distant stabilizers may be strongly correlated.
Finally, among the architectures we explored, the transformer encoder proved to be the most stable when scaling to larger system sizes, exhibiting smoother convergence curves and reduced sensitivity to hyperparameter choices.

Another architectural choice is the use of residual connections, which help preserve gradient flow through multiple layers -- crucial for the deeper networks used for larger system sizes. 
We also experimented with pre-layer normalization (pre-LN) as an alternative to the standard post-layer normalization approach, but observed no significant performance advantages.
Similarly, we tested alternative activation functions including ReLU and tanh in the feed-forward networks, but found that GELU provided smoother convergence and improved overall training stability. 
Dropout regularization was also evaluated but did not yield improvements in training performance, likely due to the structured nature of the quantum state representations.

\subsection{Training process}
\label{app:rl_training}
We employ a \textit{warmup cosine decay} learning rate schedule which improves training stability, facilitates convergence, and enables soft restarts from checkpoints -- particularly at large system sizes ($N=64,128$) where training required multiple checkpoint-based restarts. 

The schedule starts with a linear warmup for $t < T_{\text{warmup}}$, defined as
\begin{equation*}
    \eta(t) = \eta_{\text{init}} + (\eta_{\text{max}} - \eta_{\text{init}}) \frac{t}{T_{\text{warmup}}},
\end{equation*}
where $\eta_{\text{init}} = 0.01 \times$ \texttt{learning\_rate}, $\eta_{\text{max}} = $ \texttt{learning\_rate}, and $T_{\text{warmup}} = 0.02 \times$ \texttt{num\_timesteps}.

After the warmup phase, the learning rate follows a cosine decay:
\begin{equation*}
    \eta(t) = \eta_{\text{min}} + \frac{1}{2}(\eta_{\text{max}} - \eta_{\text{min}}) 
    \left[1 + \cos\left(\pi \frac{t - T_{\text{warmup}}}{T_{\text{decay}}}\right)\right],
\end{equation*}
where $\eta_{\text{min}} =$ \texttt{alpha}, and $T_{\text{decay}} = 0.98 \times$ \texttt{num\_timesteps}.

Our implementation uses JAX's \texttt{shard\_map} for distributed execution across multiple devices.
The environment rollouts -- i.e., parallel simulations used to collect trajectories -- are distributed across all available GPUs using global mesh parallelization. 
The total number of environments running in parallel (\texttt{num\_envs}) is evenly divided among the GPUs, so that each GPU evolves \texttt{num\_envs}$/$ num\_gpus environments independently.
The gradient computation and parameter updates are synchronized across all devices, and the training states are replicated across devices to ensure consistency.

To monitor convergence, the training process periodically runs evaluation episodes. A total of \texttt{num\_evals} evaluations are performed during training, each consisting of \texttt{num\_eval\_envs} environments unrolled for \texttt{eval\_episode\_length} timesteps.

Training stability varies significantly with both system size $N$ and the bias parameter $p$. 
For high bias values (e.g., $p = 0.15, 0.2$), training is stable and converges rapidly across all system sizes. 
In contrast, for low bias ($p = 0.05, 0.1$), the increased stochasticity leads to unstable training, necessitating careful hyperparameter tuning. 
Additionally, in this regime, systems take longer to reach a steady state, necessitating longer episodes and resulting in increased overall training time.

Larger systems ($N \geq 64$) present additional challenges due to the need for deeper networks and longer episodes, which makes training more unstable.
This requires more conservative learning rates ($\sim 10^{-5}$) and multiple checkpoint restarts; the learning rate schedule enables smooth continuation across restarts and avoids abrupt policy changes that could lead to performance collapse.

Important was the use of a high discount factor, $\gamma = 0.99$, which promotes long-horizon planning. This encourages the agent to account for delayed effects -- essential for discovering effective disentanglement strategies.

\begin{figure}[h]
    \centering
    \includegraphics[width=\linewidth]{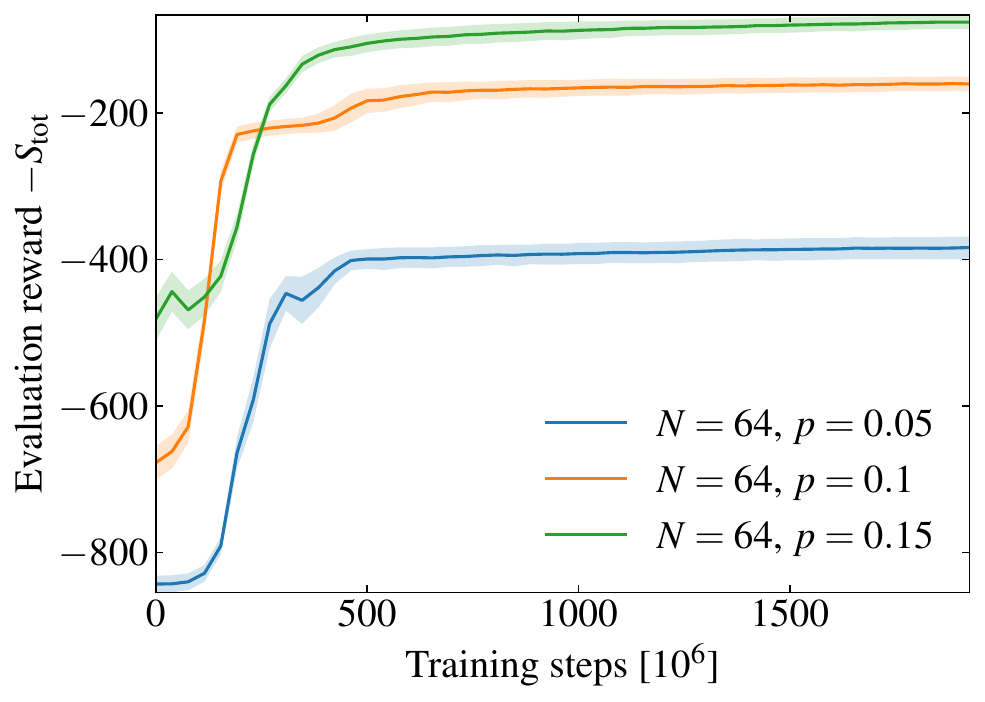}
    \caption{\textbf{Learning curves for representative models.} 
    The solid line shows the mean evaluation reward, and the shaded area indicates the standard deviation across evaluation episodes.}
    \label{fig:rl_learning_curves}
\end{figure}

To illustrate convergence and stability across parameters, we report the evaluation rewards measured throughout training (Fig.~\ref{fig:rl_learning_curves}) for three selected sets of parameters: $N=64, p=0.05,0.1,0.15$. Each curve shows the mean episode reward as a function of training steps, with a shaded region denoting one standard deviation across evaluation runs. 

\subsection{Computational resources}
\label{app:comp_resources}

All trainings are conducted on a high-performance computing system with the following specifications:
\begin{itemize}
    \item Intel Xeon Platinum 8360Y CPUs (36 cores at 2.40GHz, dual CPU per node)
    \item 72 execution nodes with 1 TB RAM and 4 Nvidia A100-80GB GPUs each
\end{itemize}

Resource allocation scaled with system size $N$:
\begin{itemize}
    \item $N = 16, 32$: 1 full node (4 A100 GPUs, 1 TB RAM)
    \item $N = 64$: 2 nodes (8 A100 GPUs, 2 TB RAM)  
    \item $N = 128$: 4 nodes (16 A100 GPUs, 4 TB RAM)
\end{itemize}

The computational resource requirements approximately scale linearly with system size. 
However, the dominant bottleneck is the transient time required for the physical dynamics to reach the steady state (see App.~\ref{app:transient_times}). This necessitates longer episode lengths and more training steps for larger systems, leading to significantly increased wall-clock training times despite the linear increase in hardware allocation.

Since all computations are performed on GPUs, CPU resources were not a limiting factor. The distributed training across multiple GPUs was handled automatically by JAX's parallelization primitives, allowing efficient scaling to larger system sizes.

\subsection{Hyperparameters}
\label{app:rl_hyperparameters}
Table~\ref{tab:hyperparams} summarizes the hyperparameters used for training across system sizes $N \in \{16, 32, 64, 128\}$. While common defaults were used when possible, some fine-tuning was required for each model. The complete list of hyperparameter configurations is available in the Zenodo repository~\cite{qtetrix_repo}.
Let us note that our aim is not to optimize these hyperparameters for peak performance. Rather, the primary objective of this work is to uncover qualitatively distinct strategies and to gain insight into the underlying physical mechanisms.

\begin{table*}[t]
\centering
\renewcommand{\arraystretch}{1.1}
\setlength{\tabcolsep}{6pt}
\begin{tabular}{l@{\hspace{12pt}}p{0.15\textwidth}@{\hspace{15pt}}p{0.15\textwidth}@{\hspace{12pt}}p{0.15\textwidth}@{\hspace{12pt}}p{0.15\textwidth}}
\toprule
\textbf{Parameter} & \textbf{$N=16$} & \textbf{$N=32$} & \textbf{$N=64$} & \textbf{$N=128$} \\

\midrule

\texttt{policy\_num\_heads} & 4, 8 & 1, 8 & 1, 8, 16 & 1, 16 \\
\thinline
\texttt{value\_num\_heads} & 4, 8 & 1, 8 & 1, 8, 16 & 1, 16 \\ 
\thinline
\texttt{policy\_feedforward\_sizes} & (256,), (512,) & (256,), (512,) & (512,), (1024,) & (1024,) \\ 
\thinline
\texttt{value\_feedforward\_sizes} & (256,), (512,) & (256,), (512,) & (512,), (1024,) & (1024,) \\ 
\texttt{policy\_num\_layers} & 1, 4 & 1, 3, 4 & 1, 5, 6 & 5, 6 \\ 
\texttt{value\_num\_layers} & 1, 4 & 1, 3, 4 & 1, 5, 6 & 5, 6 \\ 

\midrule

\texttt{num\_timesteps} & $10^9$ & $10^9$, $2 \times 10^9$ & $10^9$, $2 \times 10^9$ & $10^9$, $2 \times 10^9$* \\
\thinline
\texttt{episode\_length} & 500, 1000 & 1000 & 5000, 10000, 20000 & 10000, 25000, 50000 \\
\thinline
\texttt{eval\_episode\_length} & 1000 & 1000 & 1000 & 10000\\
\thinline
\texttt{num\_envs} & 1024, 2048 & 1024, 2048 & 1024 & 1024 \\
\thinline
\texttt{num\_eval\_envs} & 2048 & 2048 & 2048 & 2048 \\
\thinline
\texttt{unroll\_length} & 11, 35 & 10, 30 & 10, 30, 50 & 25 \\
\thinline
\texttt{num\_minibatches} & 16 & 16 & 8, 16 & 8, 16 \\
\thinline
\texttt{batch\_size} & 512, 1024 & 1024, 2048 & 1024 & 1024 \\
\thinline
\texttt{num\_updates\_per\_batch} & 5 & 1, 3, 5 & 2, 3 & 2, 3 \\
\thinline
\texttt{num\_evals} & 53 & 53 & 53 & 53\\
\thinline
\texttt{learning\_rate} & $3 \times 10^{-4}$, $5 \times 10^{-4}$ & $1 \times 10^{-5}$, $5 \times 10^{-5}$, $1 \times 10^{-4}$ & $3 \times 10^{-5}$, $1 \times 10^{-4}$ & $1 \times 10^{-5}$ \\
\thinline
\texttt{alpha} & 0, $1 \times 10^{-5}$ & 0, $1 \times 10^{-6}$ & 0, $1 \times 10^{-6}$ & 0, $1 \times 10^{-6}$ \\
\thinline
\texttt{entropy\_cost} & 0 & 0 & $1 \times 10^{-5}$, $1 \times 10^{-4}$ &  $1 \times 10^{-4}$, $1 \times 10^{-5}$ \\
\thinline
\texttt{value\_cost} & 100 & 10, 100 & 100 & 100\\
\thinline
\texttt{discounting} & 0.99 & 0.99 & 0.99 & 0.99 \\
\thinline
\texttt{reward\_scaling} & 0.01 & 0.02 & 0.02 & 0.02 \\
\thinline
\texttt{clipping\_epsilon} & 0.2 & 0.1, 0.2 & 0.1, 0.2 & 0.1, 0.2 \\
\thinline
\texttt{gae\_lambda} & 0.5 & 0.5, 0.6, 0.8, 0.95 & 0.6, 0.8, 0.95 & 0.8, 0.9 \\
\bottomrule
\end{tabular}
\caption{\textbf{Reinforcement learning hyperparameters} for each system size $N$. Multiple values indicate different configurations explored during hyperparameter tuning.}
\label{tab:hyperparams}
\end{table*}

\section{Stabilizer circuits}
\label{app:stabilizer_circuits}

This section provides a brief introduction to stabilizer circuits and related concepts used in this work. It is not intended to be a comprehensive description of Stabilizer circuits, but rather a brief summary. The more familiar people will find this useful for refreshing their memory; for the less familiar, see Refs.~\cite{gottesman2024surviving,gottesmanthesis} for an extensive and comprehensive introduction to the field.

Stabilizer circuits are important for two reasons: they can be efficiently simulated classically and implement certain quantum error correction codes, known as stabilizer codes.
The central insight of the stabilizer formalism is that a quantum state can be represented by a stabilizer group consisting of unitary operators that leave the state invariant, rather than with a vector of amplitudes. 
In some cases, this representation is more efficient. While this may appear counterintuitive at first glance, it has proven to hold true~\cite{Nielsen_Chuang_2010}.
\\

Consider a system of $n$ qubits and the \textit{Pauli strings} that act on the whole register, i.e., the $n$-fold tensor product of Pauli matrices -- in the following, we will omit tensor product signs for brevity, thus $-YXZZ := -\sigma_y \otimes \sigma_x \otimes \sigma_z \otimes \sigma_z$.
The set of all possible $4^{n+1}$ phased Pauli strings with the operation of matrix multiplication over the tensor product of Hilbert space $(\mathbb{C}^2)^{\otimes n}$ forms the \textit{Pauli group} $\mathcal{P}_n$ \cite{Gottesman_1998_Heisenberg}
\begin{equation}
	\mathcal{P}_n := \{ \alpha O_1 O_{2} \dots O_n \},
\end{equation}
where $\alpha \in \{1,-1,i,-i\}$ and $O_{j} \in \{I,X,Y,Z\}$. Since $Y = i X Z$ and $X^2 = Y^2 = Z^2 = I$, a commonly used generating set of the Pauli group is
\begin{equation}
    \{ X_1, \dots, X_n, Z_1,\dots,Z_n \},
    \label{eq:Pauli_generating_set}
\end{equation}
where $X_j := I^{\otimes(j-1)}\otimes X \otimes I^{\otimes(n-j+1)}$, that is a single qubit operator $X$ applied to the $j$th qubit in the tensor product Hilbert space $(\mathbb{C}^{2})^{\otimes n}$. Similarly for $Z_j$.

Given the notion of Pauli group, we consider its normalizer~\footnote{The normalizer of a subset $\mathcal{S}$ in a group $\mathcal{G}$ is the set of elements of $\mathcal{G}$ that leave the set $\mathcal{S}$ fixed under conjugation, i.e., $N_{\mathcal{G}}(\mathcal{S}) := \{g \in \mathcal{G} | g \mathcal{S} = \mathcal{S}g\} = \{ g\in \mathcal{G} | g \mathcal{S} g^{-1} = \mathcal{S}\}$.}
$\mathcal{N}(\mathcal{P}_{n}) = \{ U \in U(2^n) | U \mathcal{P}_n c^{\dagger} = \mathcal{P}_n \}$ in the unitary group $U(2^n)$. 
The {\it Clifford group} is defined as the normalizer of the Pauli group, neglecting the global phase $\mathcal{C}_{n}:= \mathcal{N}(\mathcal{P}_{n}) \backslash U(1)$ \cite{Gottesman_1998, Koenig_2014}. 
We disregard the global phase since we are interested in observably distinguishable Clifford gates, i.e., those that produce different measurement distributions.
While $\mathcal{C}_{n}$ is considerably smaller than the full unitary group $U(2^n)$, it still contains several operators of particular interest. The elements of the Clifford group are called \textit{Clifford gates}.

The Clifford group is generated by the Hadamard gate $H$, the phase gate $P$, and the controlled-NOT ($\mathrm{CNOT}$) gate, or $\mathrm{XOR}$ gate~\cite{gottesman2024surviving}:
\begin{align*}
    &H = \frac{1}{\sqrt{2}} \begin{bmatrix} 1 & 1 \\ 1 & -1 \end{bmatrix}, \\[2ex]
    &P = \begin{bmatrix} 1 & 0 \\ 0 & e^{i \frac{\pi}{2}} \end{bmatrix} = \begin{bmatrix} 1 & 0 \\ 0 & i \end{bmatrix}, \\[2ex]
    &\mathrm{CNOT} = \begin{bmatrix} 1 & 0 & 0 & 0 \\
                0 & 1 & 0 & 0 \\ 0 & 0 & 0 & 1 \\ 0 & 0 & 1 & 0\end{bmatrix}.
\end{align*}
Moreover, all Pauli matrices can be constructed from $P$ and $R=Z$ gates \cite{Ozols_clifford_group}. Therefore, each Pauli gate is also trivially an element of the Clifford group.\\

Given a pure quantum state $\ket{\psi}$, and a unitary operator $S$, we say that $S$ is a \textit{stabilizer} of $\ket{\psi}$, if $\ket{\psi}$ is an eigenstate with eigenvalue $+1$, i.e., $S \ket{\psi} = +\ket{\psi}$.
All the possible stabilizers of the state $\ket{\psi}$, with the multiplication operation, form the \textit{stabilizer group} $\text{Stab}(\ket{\psi})$: if $S_1, S_2 \in \mathcal{S}$ stabilize $\ket{\psi}$, then so do $S_1S_2$ as well as $S_1^{-1}$.

If $\ket{\psi} \neq \ket{\phi}$, then $\text{Stab}(\ket{\psi}) \neq \text{Stab}(\ket{\phi})$~\cite{aaronson_2004}; that is, the representation of a state in terms of its stabilizers is unique. 
Therefore, instead of tracking the time evolution of the state $\ket{\psi}$ directly, we follow the evolution of its stabilizers.
However, is this stabilizer formalism more efficient than the standard representation in terms of a vector of amplitudes?
This is a valid question, considering that, in general, the stabilizer formalism is even worse: writing down the generators of $\text{Stab}(\ket{\psi})$ takes about $2^{2n}$ parameters, instead of $2^n$ parameters needed to describe the vector of amplitudes. 
Remarkably, though, a large and interesting class of quantum states can be uniquely specified by a smaller stabilizer group, namely the intersection of $\text{Stab}(\ket{\psi})$ with the Pauli group $\mathcal{P}_n$.

In general, the evolution of a stabilizer state maps Pauli matrices to arbitrary unitaries. Under arbitrary quantum gates, tracking the evolution of the stabilizer generators quickly becomes intractable.
To avoid this, one typically restricts to Clifford gates, which preserve the Pauli group by mapping Pauli operators to other Pauli operators.
This is formalized by the following theorem~\cite{aaronson_2004}:
\begin{theorem}
	Given a $n$ qubits state $\ket{\psi}$, the following are equivalent:
	\begin{itemize}
		\itemsep0em
		\item $\ket{\psi}$ can be obtained from $\ket{0}^{\otimes n}$ by CNOT, Hadamard, and phase gates only;
		\item $\ket{\psi}$ is stabilized by exactly $2^n$ Pauli operators;
		\item $\ket{\psi}$ is uniquely determined by $\mathcal{S}(\ket{\psi}) := \text{Stab}(\ket{\psi}) \bigcap \mathcal{P}_n$, the group of Pauli operators that stabilize $\ket{\psi}$.
	\end{itemize}
	\label{th:aaronson}
\end{theorem}

Because of Theorem~\ref{th:aaronson}, any circuit with CNOT, Hadamard, phase gates, and Pauli string measurements is called a \textit{stabilizer circuit}. We call a stabilizer circuit ``unitary'' if it does not contain measurement gates. Unitary stabilizer circuits are also known as \textit{Clifford group circuits}~\cite{aaronson_2004}. 
Moreover, with a slight abuse of notation, in the following, we will refer to $\mathcal{S}$ as the stabilizer group, unless stated otherwise.

Theorem~\ref{th:aaronson} has some significant implications. 
In general, a quantum state requires $2^n$ complex amplitudes for full characterization. 
However, Theorem~\ref{th:aaronson} says that any state $\ket{\psi}$ prepared from $\ket{0}^{\otimes n}$ via a stabilizer circuit can be uniquely described in terms of $\mathcal{S}(\ket{\psi}) := \text{Stab}(\ket{\psi}) \bigcap \mathcal{P}_n$, and $|\mathcal{S}(\ket{\psi})|=2^n$.
Any finite group $\mathcal{G}$ has a generating set of size at most $\log_2|\mathcal{G}|$ elements, so $\mathcal{S}(\ket{\psi})$ can be uniquely specified by $n$ independent, commuting generators that are elements of $\mathcal{P}_n$: $\mathcal{S}(\ket{\psi}) = \langle g_1, g_2, \dots,g_n \rangle$. 
Moreover, the evolution $S \rightarrow U S U^{\dagger}$ requires tracking the generators of the stabilizer group only~\cite{Gottesman_1998_Heisenberg}.
This leads to a crucial conclusion: the wavefunction $\ket{\psi}$ of a stabilizer circuit is uniquely described by $n$ mutually commuting and independent Pauli string operators $G = \{g_1, \dots, g_\mathrm{L} \}$ such that $g_i \ket{\psi} = \ket{\psi}$.

Each generator of $\mathcal{S}$ can be encoded using $2n+1$ bits: $2$ bits for each Pauli matrix and 1 bit for the phase~\footnote{If $S \in \mathcal{S}$, then $S$ can only have a phase of $\pm1$, not $\pm i$; in the latter case $S^2 = - I \dots I$ would be in $\mathcal{S}$, but $-I$ does not stabilize anything.}. 
Since there are a total of $n$ generators, the full state can be uniquely represented using $n(2n+1)$ bits. Moreover, these bits can be efficiently updated (in polynomial time) following the application of a Clifford or measurement gate. 
Consequently, the time and space required to analyze a stabilizer circuit on classical computers is polynomial in $n$, rather than exponential.
This is the conclusion of the following theorem \cite{Gottesman_1998}.
\begin{theorem}[Gottesman-Knill]
	Any quantum computer performing only:
	\begin{itemize}
		\itemsep0em
		\item Clifford group gates,
		\item measurements of Pauli group operators, and
		\item classical control, through Clifford group operations, conditioned on the results of earlier measurements,
	\end{itemize}
	can be perfectly simulated in polynomial time on a classical computer.
\end{theorem}
The Gottesman–Knill theorem highlights the subtle power of quantum computation. It shows that some quantum computations involving highly entangled states may be efficiently simulated on classical computers.

\subsection{Simulating stabilizer circuits}
\label{app:simulating_stabilizers}
First introduced by Aaronson and Gottesman~\cite{aaronson_2004}, the Tableau algorithm represents an efficient method to simulate stabilizer circuits on classical computers.
The central component of the algorithm is the so-called \textit{tableau} that represents the $n$ stabilizer and the $n$ destabilizer generators. The construction of the tableau is straightforward. 
We label the destabilizer generators as $R_1, \dots, R_n$ and the stabilizer generators as $R_{n+1}, \dots, R_{2n}$, where $R_i$ will then represent a row in the tableau. 
Each generator is an element of the Pauli group, and it can be written as $R_i = \pm O_1 O_2 \dots O_n$, where $O_j$ are single-qubit operators acting on the $j$th spin in the register. 
Given there are four possible operators $I,Z,X,Y$, to encode each operator $O_{j}$ we need 2 classical bits $x_{ij}$ and $z_{ij}$, shown in Table~\ref{tab:operators_encoding}.
We need one additional classical bit $r_i$ to encode the phase: $r_i = 1$ for the $-$ sign and $r_i=0$ for the $+$ sign.  
\begin{table}[h]
    \centering
    \begin{tabular}{ c|cc }
        $P_{ij}$ & $x_{ij}$ & $z_{ij}$ \\
        \hline
        $I$        & 0        & 0        \\
        $Z$        & 0        & 1        \\
        $X$        & 1        & 0        \\
        $Y$        & 1        & 1        \\
    \end{tabular}
    \caption{Table showing the encoding of Pauli matrices in the tableau formalism. Here $P_{ij}$ represents the $j^{\text{th}}$ single qubit operator in the $i^{\text{th}}$ row $R_i$.}
    \label{tab:operators_encoding}
\end{table}
Therefore the tableau consists of binary variables $x_{ij}$, $z_{ij}$ and $r_i$, where $i \in \{ 1,\dots,2n\}$ and $j \in \{1,\dots,n\}$, and it is written as
\begin{equation*}
    \centering
    \begin{pmatrix}
        \begin{array}{ccc|ccc|c}
            x_{11}     & \dots  & x_{1n}     & z_{11}     & \dots  & z_{1n}     & r_1     \\
            \vdots     & \ddots & \vdots     & \vdots     & \ddots & \vdots     & \vdots  \\
            x_{n1}     & \dots  & x_{nn}     & z_{n1}     & \dots  & z_{nn}     & r_n     \\
            \hline
            x_{(n+1)1} & \dots  & x_{(n+1)n} & z_{(n+1)1} & \dots  & z_{(n+1)n} & r_{n+1} \\
            \vdots     & \ddots & \vdots     & \vdots     & \ddots & \vdots     & \vdots  \\
            x_{(2n)1}  & \dots  & x_{(2n)n}  & z_{(2n)1}  & \dots  & z_{(2n)n}  & r_{2n}  \\
        \end{array}
    \end{pmatrix}
\end{equation*}
As an example, the 2-qubit state $\ket{\psi} = \ket{00}$ is stabilized by $\mathcal{S}(\ket{\psi}) = \{+ZI, +IZ\}$, that is $R_3 = +ZI$ and $R_4 = +IZ$. It is easy to see that the destabilizer generators are $R_1 = +XI$ and $R_2=+IX$, because together with the stabilizers they generate the whole group $\mathcal{P}_2$. The tableau reads
\begin{equation*}
	\centering
	\begin{pmatrix}
		\begin{array}{cc|cc|c}
			1 & 0 & 0 & 0 & 0 \\
			0 & 1 & 0 & 0 & 0 \\
			\hline
			0 & 0 & 1 & 0 & 0 \\
			0 & 0 & 0 & 1 & 0 \\
		\end{array}
	\end{pmatrix}
	\label{eq:identity-tableau}
\end{equation*}
In this work, we use a reduced tableau representation, omitting both the destabilizers and the phase; this results in an $n \times 2n$ matrix.
Note that the tableau representation is not unique. There is a large gauge freedom in defining the stabilizers, as one can freely redefine them by multiplying together different stabilizers~\cite{nahum_2017}. For example, if a state is stabilized by $\{XI, IZ\}$, then it is also stabilized by $\{XZ, IZ\}$.

In the Tableau formalism, the gates are written based on how they transform the basis elements. For example, in the case of a 2-qubit gate, we can fully characterize a gate by specifying how it maps the following 4 Pauli strings to themselves $XI, IX, ZI, IZ$ or $X_1, X_2, Z_1, Z_2$. Therefore, the representation of a 2-qubit gate is a $4 \times 4$ matrix. For example, the CNOT gate is represented as the following binary
\begin{table}[h]
    \centering
    \begin{tabular}{ c|cccc }
              & $X_1$ & $X_2$ & $Z_1$ & $Z_2$ \\
        \hline
        $X_1$ &  1    &   0   &   0   &   0   \\
        $X_2$ &  1    &   1   &   0   &   0   \\
        $Z_1$ &  0    &   0   &   1   &   1   \\
        $Z_2$ &  0    &   0   &   0   &   1   \\
    \end{tabular}
    \caption{Tableau representation of the CNOT gate.}
    \label{tab:cnot_example}
\end{table}
Given this representation, the action of a gate is a simple matrix multiplication modulo 2. There are two possible approaches: 
\begin{itemize}
    \item perform the matrix multiplication on the indices that describe the sub-system,
    \item embed the gate in the full system and matrix multiply the tableau times the gate.
\end{itemize}

\section{Disentangling Clifford gates}
\label{app:disentangling_cliffords}
In this section, we present the methodology used to find optimal two-qubit disentangling Clifford gates. Specifically, given a stabilizer state and a pair of neighboring qubits, we seek the Clifford gate that maximally disentangles that pair.
Restricting to stabilizer circuits simplifies the problem significantly: the circuit dynamics can only access a finite set of states, a direct consequence of the Clifford group having a finite set of gates.
Working with stabilizer states, we look at the representation in terms of stabilizers. 
In a two-qubit subsystem, there can be at most four independent stabilizers: the stabilizer group of a two-qubit system is generated by $\{ IX, XI, ZI, IZ \}$, where for brevity we use the notation $XI := \sigma_x \otimes \id_2$.

To classify different states, we consider the Tableau representation in the \textit{clipped gauge}~\cite{nahum_2017,li_2019}. 
As mentioned in the previous section, the Tableau representation of a stabilizer state $\ket{\psi}$ is not unique. The $n$ generators $G$ can be multiplied by one another without altering the represented wavefunction. The clipped gauge partially removes this degeneracy while also highlighting the entanglement entropy structure. To introduce the clipped gauge, we look at the stabilizers and their support. Being Pauli strings, we define left and right endpoints as:
\begin{align}
    \texttt{l}(g) = \min \{ \text{i: g acts non trivially on site i} \}, \\
    \texttt{r}(g) = \max \{ \text{i: g acts non trivially on site i} \}.
\end{align}
For any stabilizer state, one can choose $G$ such that there are exactly two stabilizer endpoints on each site:
\begin{equation}
    n_\texttt{l}(i) + n_\texttt{r}(i) = 2, \quad \text{for all $i$},
    \label{eq:clipped_condition}
\end{equation}
where $n_\texttt{l}(i)$ ($n_\texttt{r}(i)$) is the number of left (right) endpoints on site $i$.
If Eq.~\eqref{eq:clipped_condition} is satisfied, $G$, hence the Tableau, is in the clipped gauge.
In other words, each stabilizer is "localized" in a minimal range, meaning its nonzero entries appear within a well-defined segment.
When $G$ is in the clipped gauge, the entanglement entropy of a contiguous subregion $A$ equals half the number of stabilizers crossing either its left or right boundary.
\\

We construct a lookup table of disentangling gates by enumerating all possible two-qubit stabilizer states, embedded within a larger chain and expressed in the clipped gauge. For each configuration, we determine the corresponding disentangling Clifford gate and store it.
This creates a one-to-one mapping between two-qubit states and disentangling gates.
This precomputation enables a highly efficient disentangling algorithm, as the lookup table removes the need for additional computation -- all necessary gate searches are performed in advance.

There are a total of 21 distinct stabilizer structures, as shown in Figure~\ref{fig:stabilizers_classes}, depicted using a diagrammatic representation. 
The support of the stabilizers is illustrated with a line: different colors represent different stabilizers. 
In the figure, we limit the depiction to stabilizers that have either the left or right endpoint within the two-qubit system; other stabilizers that act non-trivially on the qubits may exist, but they are redundant for representing the state and unnecessary for the disentangling procedure. 
Applying a gate on two qubits can only alter the endpoints on those qubits, not on others.

\input{tikz_diagrams/stabilizer_diagrams}

Not all of the listed arrangements can be disentangled. Some are already minimally entangled: 2.1, 3.1, 3.1*, 3.5, 4.1, 4.2, 4.2*, 4.4, and 4.4*. 
In these cases, we leave the state unchanged.
For other cases, the system can be disentangled by moving the edges of the stabilizers, as follows: 2.2 $\rightarrow$ 2.1, 3.3 $\rightarrow$ 3.1*, 3.3* $\rightarrow$ 3.1, 3.4 $\rightarrow$ 3.1, 3.4* $\rightarrow$ 3.1*, 3.6 $\rightarrow$ 3.5, 4.3 $\rightarrow$ 4.2, 4.3* $\rightarrow$ 4.2*, 4.5 $\rightarrow$ 4.1, 4.6 $\rightarrow$ 4.1. 
For these cases, we iterate through $\mathcal{C}_2$ to find the Clifford gate that performs the necessary transformation. 
While each of the mappings above preserves the clipped gauge, this is not true in general—applying an arbitrary Clifford gate typically disturbs the gauge. As a result, the clipped gauge must be restored after the disentangling gate is applied.

Two arrangements were left out: 3.2 and 3.2*. Both cannot be further disentangled using the action of a two-qubit gate. However, both can be mapped to 3.5. 
These cases are considered special as they do not immediately reduce entanglement, but the full state they lead to is easier to disentangle. 
For a generic system, there is always the possibility to act on a state in a way that decreases entanglement. As such, a greedy strategy will not use these states. However, these special cases may be significant in a more complex entanglement-reducing strategy.

Another approach is presented in~\cite{yepes_2024}, where they rely on 9 total gates to reduce entanglement of a two-qubit system.

\subsection{Efficient clipping algorithm}
\label{app:clipping_algorithm}
Given a stabilizer state $\ket{\psi}$, mapping it to disentangling gates requires the tableau to be in clipped gauge. This is done at each time step of the dynamics, and an efficient clipping algorithm is essential to keep computation time manageable.

A method for gauge fixing is described in~\cite{li_2019}. It involves two successive Gaussian eliminations on the tableau, requiring $O(N^2)$ row operations and conditional branching. These operations can be computationally expensive.
Our implementation achieves the same result using efficient bitwise operations and parallel reductions. This makes it well-suited for large matrices and GPU execution.

The algorithm has two main stages:
\begin{itemize}
    \item \textbf{Pivot selection:} 
    Each column is processed in sequence. The pivot is chosen as the shortest unprocessed stabilizer with an endpoint on that column. This selection uses a \texttt{clip\_map}, which tracks the left and right endpoints of each stabilizer.

    \item \textbf{Row reduction:} 
    Once the pivot is selected, other relevant rows are reduced using bitwise XOR. Since tableaux are binary matrices, XOR naturally implements stabilizer multiplication. The \texttt{clip\_map} is updated dynamically to reflect changes in row supports.
\end{itemize}

The algorithm begins by fixing the left endpoints of all stabilizers. It processes columns from left to right, selecting pivots and applying row reductions. Then it fixes the right endpoints by repeating the process from right to left.
Choosing the correct pivot efficiently is a central challenge. Instead of scanning all rows, the algorithm uses the \texttt{clip\_map}, an $N \times 2$ matrix storing the left and right endpoints of each row. This compact representation allows fast access to row supports.

The pivot is selected based on the following rules:
\begin{enumerate} 
    \item The row must not have been processed. 
    \item It must have an endpoint on the current column (left or right, depending on direction). 
    \item Among these, the row with the smallest support is chosen. 
\end{enumerate}

Once the pivot is found, it is used to eliminate all other rows with an endpoint on the same column. This proceeds as follows:
\begin{enumerate} 
    \item Identify the pivot row. 
    \item Create a boolean mask to select rows that need updating. 
    \item Apply a single bitwise XOR operation to update all selected rows in parallel. 
\end{enumerate}
This parallelized approach significantly reduces computation time.

Finally, the clipped gauge also enables direct computation of entanglement. The number of stabilizers crossing a given cut determines the entanglement across that cut.

\subsubsection{Performance considerations}
This algorithm is significantly more efficient than traditional Gaussian elimination for stabilizer tableaux due to the following factors:
\begin{itemize}
    \item \textbf{Parallelization:} Operations are applied to multiple rows simultaneously using JAX's vectorized functions.
    \item \textbf{Minimal branching:} The implementation avoids conditionals inside loops, which improves performance on modern hardware.
    \item \textbf{Bitwise operations:} XOR-based row updates are computationally cheaper than arithmetic operations.
\end{itemize}
These optimizations make the algorithm well-suited for large-scale stabilizer simulations.

\end{appendix}

\bibliography{references}

\end{document}

%% file: tikz_diagrams/bipartite_graph.tex
\definecolor{csystem}{HTML}{5dade2}
\definecolor{cagent}{HTML}{e74c3c}

\begin{figure}
    \centering

    \begin{tikzpicture}[scale=1.1]
        \coordinate (system) at (0,0);
        \coordinate (agent) at (2,0);
        \coordinate (composite) at (5.5,0);
        \def\u{0.8cm};
        \coordinate (v) at (0,\u);
        \coordinate (h) at (\u,0);
        \def\r{0.08cm};
        
        \path[csystem,thick] ($(system) + (v)$) edge[bend right=40] ($(system) - (v)$);
        \path[csystem,thick] ($(system) + (v)$) edge[bend left=40] ($(system) - (v)$);
        \fill[csystem] ($(system) + (v)$) circle (\r);
        \fill[csystem] ($(system) - (v)$) circle (\r);
    
        \node[left] at ($(system) + (v) - (0.2,0)$) {$y=1$};
        \node[left] at ($(system) - (v) - (0.2,0)$) {$y=0$};
        
        \node[scale=1] at ($0.5*(system) + 0.35*(agent)$) {$\times$};
            
        \draw[cagent,thick,dashed] ($(agent) - (h)$) -- ($(agent) + (h)$);
        \fill[cagent] ($(agent) + (h)$) circle (\r);
        \fill[cagent] ($(agent) - (h)$) circle (\r);
        \node[below] at ($(agent) - (h) - (0,0.2)$) {$x=0$};
        \node[below] at ($(agent) + (h) - (0,0.2)$) {$x=1$};
            
        \draw[-{Stealth[scale=1.]}, thick] ($(agent) + 1.3*(\u,0)$) -- ($(agent) + 1.8*(\u,0)$);
        
        \def\zz{1.5};
        \draw[-{Stealth}] ($(composite) - \zz*(h) - \zz*(v)$) -- ($(composite) - \zz*(h) + \zz*(v)$);
        \draw[-{Stealth}] ($(composite) - \zz*(h) - \zz*(v)$) -- ($(composite) + \zz*(h) - \zz*(v)$);
    
        \node[below] at ($(composite) - (h) - \zz*(v) - (0,0.1)$) {$0$};
        \draw[black] ($(composite) - (h) - \zz*(v) - (0,0.05)$) -- ($(composite) - (h) - \zz*(v) + (0,0.05)$);
        \node[below] at ($(composite) + (h) - \zz*(v) - (0,0.1)$) {$1$};
        \node[below] at ($(composite) + (h) - \zz*(v) - (0,0.1)+(0.4,0)$) {$x$};
        \draw[black] ($(composite) + (h) - \zz*(v) - (0,0.05)$) -- ($(composite) + (h) - \zz*(v) + (0,0.05)$);
        \node[left] at ($(composite) - \zz*(h) - (v) - (0.1,0)$) {$0$};
        \draw[black] ($(composite) - \zz*(h) - (v) - (0.05,0)$) -- ($(composite) - \zz*(h) - (v) + (0.05,0)$);
        \node[left] at ($(composite) - \zz*(h) + (v) - (0.1,0)$) {$1$};
        \node[left] at ($(composite) - \zz*(h) + (v) - (0.1,0) + (0,0.4)$) {$y$};
        \draw[black] ($(composite) - \zz*(h) + (v) - (0.05,0)$) -- ($(composite) - \zz*(h) + (v) + (0.05,0)$);
    
        \path[csystem,thick] ($(composite) -(h) +(v)$) edge[bend right=40] ($(composite) -(h) -(v)$);
        \path[csystem,thick] ($(composite) -(h) +(v)$) edge[bend left=40] ($(composite) -(h) -(v)$);
        \path[csystem,thick] ($(composite) +(h) +(v)$) edge[bend right=40] ($(composite) +(h) -(v)$);
        \path[csystem,thick] ($(composite) +(h) +(v)$) edge[bend left=40] ($(composite) +(h) -(v)$);

        \draw[-{Stealth[scale=0.6]}, thick, black] ($(composite) -(h) +(0.2,0.35)$) arc (30:345:0.2);
        \draw[-{Stealth[scale=0.6]}, thick, black] ($(composite) -(h) +(0.2,-0.35)$) arc (-30:-345:0.2);
        \draw[-{Stealth[scale=1]}, thick, black] ($(composite) +(-0.45,0.35)$) arc (140:-140:0.5);
        
        \draw[cagent,thick,dashed] ($(composite) +(v) -(h)$) -- ($(composite) +(v) +(h)$);
        \draw[cagent,thick,dashed] ($(composite) -(v) -(h)$) -- ($(composite) -(v) +(h)$);
    
        \fill[shading = axis,rectangle,left color=csystem,right color=cagent,shading angle=135] ($(composite) +(v) -(h)$) circle (\r);
        \fill[shading = axis,rectangle,left color=csystem,right color=cagent,shading angle=45] ($(composite) -(v) -(h)$) circle (\r);
        \fill[shading = axis,rectangle,left color=csystem,right color=cagent,shading angle=-135] ($(composite) +(v) +(h)$) circle (\r);
        \fill[shading = axis,rectangle,left color=csystem,right color=cagent,shading angle=-45] ($(composite) -(v) +(h)$) circle (\r);
    
    \end{tikzpicture}
    \label{fig:bipartite_graph}
    \caption{\textbf{Schematic of the composite bipartite system.} 
    The physical system with states $y=0,1$ (light blue dots) connected by transitions (solid lines) is coupled to the controlling agent with states $x=0,1$ (no knowledge and knowledge states, respectively).
    The resulting composite system exhibits four distinct states: $(0,0),(0,1),(1,0),(1,1)$. The circles with arrowheads indicate the direction of probability flow, revealing the cyclic currents that violate detailed balance and the nonequilibrium nature of the state.}
\end{figure}

%% file: tikz_diagrams/stabilizer_diagrams.tex
\newcommand{\qubitradius}{0.2}  
\newcommand{\qubitdist}{0.6}    
\newcommand{\loopheight}{0.5}   

\newcommand{\stabilizerdiagram}[5]{ 
    \begin{tikzpicture}
        \coordinate (Q1) at (-\qubitdist,0);
        \coordinate (Q2) at (\qubitdist,0);

        \shade[ball color=lightgray] (Q1) circle (\qubitradius);
        \shade[ball color=lightgray] (Q2) circle (\qubitradius);

        \node[] at (0,-\qubitradius-0.2) {#1};

        \ifx#1\empty\else \draw[thick, color=red] #2; \fi 
        \ifx#2\empty\else \draw[thick, color=Blue] #3; \fi 
        \ifx#3\empty\else \draw[thick, color=BurntOrange] #4; \fi
        \ifx#4\empty\else \draw[thick, color=Green] #5; \fi
    \end{tikzpicture}
}

\newcommand{\singlesite}[1]{
    (#1) ++(0,\qubitradius) to[out=120, in=180] 
    ++(0,\loopheight) to[out=0, in=60] 
    ++(0,-\loopheight)
}
\newcommand{\twosites}[1]{
    (Q1) ++(0,\qubitradius) to[out=100, in=180] 
    ++(\qubitdist, \loopheight+#1) to[out=0, in=80] 
    ++(\qubitdist,-\loopheight-#1)
}
\newcommand{\rightshort}[2]{
    (#1) ++(0,\qubitradius) to[out=90, in=180] 
    ++(\qubitdist, \loopheight+#2)
}
\newcommand{\rightlong}[2]{
    (#1) ++(0,\qubitradius) to[out=90, in=180] 
    ++(\qubitdist, \loopheight+#2) to[out=180, in=180]
    ++(2*\qubitdist,0)
}
\newcommand{\leftshort}[2]{
    (#1) ++(0,\qubitradius) to[out=90, in=0] 
    ++(-\qubitdist, \loopheight+#2)
}
\newcommand{\leftlong}[2]{
    (#1) ++(0,\qubitradius) to[out=90, in=0] 
    ++(-\qubitdist, \loopheight+#2) to[out=0, in=0]
    ++(-2*\qubitdist,0)
}

\begin{figure*}[t]
    \centering
    \stabilizerdiagram{2.1}{
        \singlesite{Q1}
    }{
        \singlesite{Q2}
    }{}{}
    \hspace{2cm}
    \stabilizerdiagram{2.2}{
        \twosites{0}
    }{
        \twosites{0.2}
    }{}{}

    \vspace{7pt}
    \hrule
    \vspace{7pt}
    \parbox{\textwidth}{
        \stabilizerdiagram{3.1}{
            \singlesite{Q1}
        }{
            \rightshort{Q2}{0}
        }{
            \rightshort{Q2}{0.2}
        }{}
        \hspace{0.9cm}
        \stabilizerdiagram{3.1*}{
            \singlesite{Q2}
        }{
            \leftshort{Q1}{0}
        }{
            \leftshort{Q1}{0.2}
        }{}
        \hspace{0.9cm}
        \stabilizerdiagram{3.2}{
            \singlesite{Q1}
        }{
            \rightshort{Q2}{0.2}
        }{
            \leftlong{Q2}{0.2}
        }{}
        \hspace{0.9cm}
        \stabilizerdiagram{3.2*}{
            \singlesite{Q2}
        }{
            \leftshort{Q1}{0.2}
        }{
            \rightlong{Q1}{0.2}
        }{}
        \hspace{0.9cm}
        \stabilizerdiagram{3.3}{
            \singlesite{Q1}
        }{
            \leftlong{Q2}{0.2}
        }{
            \leftlong{Q2}{0.4}
        }{}
    }
    \parbox{\textwidth}{
        \stabilizerdiagram{3.3*}{
            \singlesite{Q2}
        }{
            \rightlong{Q1}{0.2}
        }{
            \rightlong{Q1}{0.4}
        }{}
        \hspace{0.9cm}
        \stabilizerdiagram{3.4}{
            \twosites{0}
        }{
            \rightlong{Q1}{0.2}
        }{
            \rightshort{Q2}{0}
        }{}
        \hspace{0.9cm}
        \stabilizerdiagram{3.4*}{
            \twosites{0}
        }{
            \leftlong{Q2}{0.2}
        }{
            \leftshort{Q1}{0}
        }{}
        \hspace{0.9cm}
        \stabilizerdiagram{3.5}{
            \leftshort{Q1}{0}
        }{
            \twosites{0}
        }{
            \rightshort{Q2}{0}
        }{}
        \hspace{0.9cm}
        \stabilizerdiagram{3.6}{
            \twosites{0}
        }{
            \rightlong{Q1}{0.2}
        }{
            \leftlong{Q2}{0.4}
        }{}
    }
    \vspace{9pt}
    \hrule
    \vspace{7pt}
    \parbox{\textwidth}{
        \stabilizerdiagram{4.1}{
            \leftshort{Q1}{0}
        }{
            \leftshort{Q1}{0.2}
        }{
            \rightshort{Q2}{0}
        }{
            \rightshort{Q2}{0.2}
        }
        \hspace{0.6cm}
        \stabilizerdiagram{4.2}{
            \leftshort{Q1}{0}
        }{
            \leftshort{Q1}{0.2}
        }{
            \leftlong{Q2}{0.4}
        }{
            \rightshort{Q2}{0}
        }
        \hspace{0.6cm}
        \stabilizerdiagram{4.2*}{
            \rightshort{Q2}{0}
        }{
            \rightshort{Q2}{0.2}
        }{
            \rightlong{Q1}{0.4}
        }{
            \leftshort{Q1}{0}
        }
        \hspace{0.6cm}
        \stabilizerdiagram{4.3}{
            \leftshort{Q1}{0}
        }{
            \rightlong{Q1}{0}
        }{
            \leftlong{Q2}{0.2}
        }{
            \leftlong{Q2}{0.4}
        }
        \hspace{0.6cm}
        \stabilizerdiagram{4.3*}{
            \rightshort{Q2}{0}
        }{
            \leftlong{Q2}{0}
        }{
            \rightlong{Q1}{0.2}
        }{
            \rightlong{Q1}{0.4}
        }
    }

    \parbox{\textwidth}{
        \stabilizerdiagram{4.4}{
            \leftshort{Q1}{0}
        }{
            \leftshort{Q1}{0.2}
        }{
            \leftlong{Q2}{0.4}
        }{
            \leftlong{Q2}{0.6}
        }
        \hspace{1.2cm}
        \stabilizerdiagram{4.4*}{
            \rightshort{Q2}{0}
        }{
            \rightshort{Q2}{0.2}
        }{
            \rightlong{Q1}{0.4}
        }{
            \rightlong{Q1}{0.6}
        }
        \hspace{1.2cm}
        \stabilizerdiagram{4.5}{
            \leftshort{Q1}{0}
        }{
            \rightlong{Q1}{0.2}
        }{
            \rightshort{Q2}{0}
        }{
            \leftlong{Q2}{0.4}
        }
        \hspace{1.2cm}
        \stabilizerdiagram{4.6}{
            \rightlong{Q1}{0}
        }{
            \rightlong{Q1}{0.2}
        }{
            \leftlong{Q2}{0.4}
        }{
            \leftlong{Q2}{0.6}
        }
    }
    \label{fig:stabilizers_classes}
    \caption{\textbf{All 21 possible local stabilizer structures in a diagrammatic representation} illustrating the support of stabilizers within a two-qubit subsystem. Different colors distinguish different stabilizers.
    A stabilizer can have support on a single qubit (e.g., diagram 2.1) or it can connect two qubits with either one or both edges in the subsystem. Additionally, a stabilizer may act non-trivially within the subsystem without visible edges. However, we exclude this case, as such a qubit can always be expressed as a linear combination of stabilizers with edges inside the subsystem, given that the stabilizer group is abelian.
    The numbering scheme is arbitrary but follows a consistent pattern: $X.Y$ where $X$ represents the number of stabilizers and $*$ labels symmetrical arrangements.
    }
\end{figure*}